\documentclass[superscriptaddress,aps,pra,twocolumn,showpacs,nofootinbib,longbibliography,notitlepage]{revtex4-2}
\usepackage{etex}
\usepackage{amsmath,amssymb,amsthm}
\usepackage[colorlinks=true,citecolor=blue,urlcolor=blue]{hyperref}
\usepackage[pdftex]{graphicx}
\usepackage{times,txfonts}
\usepackage{braket}
\usepackage{color}
\usepackage{natbib}
\usepackage{amsmath,blkarray}
\usepackage{mathtools}
\usepackage{latexsym}
\usepackage{tabularx, booktabs}
\usepackage{graphics,epstopdf}
\usepackage{graphicx}
\usepackage{float}
\usepackage{graphicx}
\usepackage{amsfonts}
\usepackage{subcaption}
\usepackage{color,soul}

\newcommand{\be}{\begin{equation}}
	\newcommand{\ee}{\end{equation}}
\newcommand{\ba}{\begin{eqnarray}}
	\newcommand{\ea}{\end{eqnarray}}

\begin{document}
\title{Generalized parity-oblivious communication games powered by quantum preparation contextuality}
\author{Prabuddha Roy}
	\email{prabuddhar@iiserbpr.ac.in}
 \affiliation{Department of Mathematical Sciences, Indian Institute of Science, Education $\&$ Research (IISER) Berhampur,\\ Transit Campus,
Govt. ITI, NH 59, Berhampur 760 010, Ganjam, Odisha, India}
	\author{A. K. Pan}
	\email{akp@phy.iith.ac.in}
	 \affiliation{Department of Physics, Indian Institute of Technology Hyderabad, Telengana-502284, India }
\begin{abstract}

The parity-oblivious random-access-code (PORAC) is a class of communication games involving a sender (Alice) and a receiver (Bob). In such games, Alice's amount of communication to Bob is constraint by the parity-oblivious (PO) conditions, so that the parity information of her inputs remains oblivious to Bob. The PO condition in an operational theory is equivalently represented in an ontological model that satisfies the preparation noncontextuality. In this paper, we provide a nontrivial generalization of the existing two-level PORAC and derive the winning probability of the game in the preparation noncontextual ontological model. We demonstrate that the quantum theory outperforms the preparation noncontextual model by predicting higher winning probability in our generalized PORAC. 
\end{abstract}
\maketitle
\section{Introduction}	One of the major objectives of quantum information theory is to find novel tasks for which quantum theory showcases its supremacy over the classical theory in terms of communication, computational efficiency, and secrecy.  Communication games  \cite{Wiesner,Nayak,Wolf,Brassard,Buhrman,zu2017,dori2021,vaisakh2021,gupta2023} are such information processing tasks involving two or more parties who collaborate to perform a given task with the highest possible efficiency with bounded amount of  communication. 
	
Consider a scenario where two parties, Alice and Bob, collaborate to perform a common task through one-way communication. In quantum theory, such a game can be played in entanglement-assisted scenario\cite{Bennett,Gavinsky,Laplante,Tavakoli2018,Muhammad,tava16,hameedi17a,pan20,Banik,hameedi17b,ghorai,asmita,mercin2017,pan2021pra,pive2022,xiao2023,roy2023njp}  prepare-and-measure scenario \cite{Galvao2001,acin,pow11,nico13,him17,armin18,kartik,tava2020,Guerin,Smania,ambainis2009,Tavakoli2015,Hayashi,spek09,Chailloux,saha19a,saha19b,Ambainis2019,pauwels2022,abhu2023}, and in hybrid scenario, i.e., entanglement-assisted prepare-measure scenario\cite{armin2021}. To exhibit the nontrivial quantum advantage in such a game, well-defined constraints on the communication have been imposed, leading to two broad classes of games. i) The communication game with bounded classical or quantum dimension \cite{Brassard,Wolf,Buhrman,acin,pan20}. ii) The oblivious communication game \cite{spek09,hameedi17b,saha19a,saha19b,Chailloux,Banik,Ambainis2019,asmita,li,abhu2023,pan2021pra}, where one puts no restriction on the amount of communication or dimension but imposes constraints so that the information of a particular property of the inputs is not transmitted due to the communication. However, there exist other communication games with energy constraints \cite{him17} and information content constraints \cite{tava2020} to the inputs. It has already been demonstrated in a plethora of works that the tasks played with quantum resources can outperform its classical counterpart\cite{Bennett,Gavinsky,Laplante,Tavakoli2018,Muhammad,tava16,hameedi17a,pan20,Banik,hameedi17b,ghorai,asmita,Galvao2001,acin,pow11,nico13,him17,armin18,kartik,tava2020,Guerin,Smania,ambainis2009,Tavakoli2015,Hayashi,spek09,Chailloux,saha19a,saha19b,Ambainis2019,gupta2023}.
	
We focus here on a specific two-party communication game - the PORAC game \cite{spek09,hameedi17b,saha19a,saha19b,Chailloux,Banik,Ambainis2019,asmita}.  In a  two-level $n$-bit PORAC, Alice receives length-$n$ bit strings as inputs, randomly sampled from $\{0,1\}^n$. Bob randomly receives index $y \in [n]$. They win the game when Bob outputs the  $y^{\text{th}}$ bit of Alice's input, i.e., $b=x_{y}$. They collaborate to optimize the winning probability of the game. Alice may communicate some bits of information to Bob to help him. However,  there is a constraint so that  Alice's communication must not allow Bob to extract the parity information of her inputs. Such a game was first put forward by Spekkens \emph{et. al.,} \cite{spek09} for two-level PORAC and later generalized for an arbittrary $d$-level case in \cite{Tavakoli2015}. It has been argued \cite{spek09, Tavakoli2015,Chailloux,saha19b} that the satisfaction of the PO condition in quantum theory can be equivalently represented  in a preparation noncontextual ontological model of quantum theory. It was shown \cite{spek09,Chailloux,ghorai,pan20} that in two-level $n$-bit PORAC the optimal quantum success probability exceeds the success probability  for a preparation noncontextual model. Therefore, any quantum advantage in PORAC reveals the preparation contextuality.

We note here that  in Spekkens \emph{et. al.,} \cite{spek09} (followed by others \cite{hameedi17b,saha19a,saha19b,ghorai,Chailloux,Banik,Ambainis2019}), the PO conditions in the two-level $n$-bit PORAC game are defined with reference to a specific parity-set, say, $\mathbb{G}_{n,2}$. The details will be specified soon. However, there is a scope for generalizing the parity-set, and if so, then each of such sets can lead to a new PORAC game. 
 
In this paper, we introduce a generalized parity-set $\mathbb{G}_{n, g_n}$ where $g_{n} \in \{2, n-1\}$ in the two-level $n$-bit PORAC game. For $g_{n}=2$, we recover the standard scenario of Spekkens \emph{et. al.} \cite{spek09}. For a given $n$, we then have $n-1$ numbers of different PORAC games. We derive the maximum success probabilities of our generalized PORAC in a preparation noncontextual model. Further, we demonstrate the quantum supremacy over the preparation noncontextual model. Note that for $2$-bit PORAC there is only one game and hence no generalization is possible.  

 We first demonstrate the quantum advantage over the preparation noncontextual model for  $3$-bit PORAC  in the prepare-and-measure scenario.  Instead of one parity-set in Spekkens \emph{et. al.} \cite{spek09}, we have two parity-sets in our $3$-bit PORAC thereby  leading to two different PORAC game. One of them is the same as in \cite{spek09}. We propose an experimental setup for testing $3$-bit generalized PORAC. We also provide a sketch of the entanglement-assisted version of our generalized $3$-bit PORAC, exhibiting the same optimal quantum advantage. We extend our study for $4$-bit generalized PORAC where we have three different games corresponding to the four different  parity sets. Further, we demonstrate the optimal quantum advantage over preparation noncontextual model for an arbitrary $n$-bit PORAC corresponding to a specific parity-set. 

 The paper is organized as follows. In Sec. II, we briefly recapitulate the essence of the two-level $n$-bit PORAC and corresponding preparation noncontextual model. In Sec. III, we provide our generalized version of two-level $n$-bit PORAC and derive the optimal success probability in a  preparation noncontextual model. In Sec. IV, we provide explicit encoding and decoding schemes to derive the optimal quantum success probability for $3$-bit PORAC game. In Sec. V. we derive the optimal quantum success probabilities for $4$-bit PORAC and further generalization for an arbitrary $n$ in Sec. VI. Finally, we summarize and discuss our work in Sec. VII.
	
\section{Preliminaries} 

 Let us first briefly summarize the notion of preparation noncontextuality in an ontological model of operational quantum theory and the standard PORAC \cite{spek09}.  
\subsection{Ontological model of an operational theory and the notion of noncontextuality}
 The modern framework of the ontological model of an operational theory was introduced in  \cite{hari,spek05}. Given a preparation procedure $P$ and a measurement procedure $M$, an operational theory provides the probability $p(k|P, M)$ of obtaining a particular outcome $k$.  If the operational theory is quantum theory, then a preparation procedure $(P)$ produces a density matrix $\rho$, and the measurement procedure $(M)$, in general, realizes positive-operator-valued measure $E_k$.  The  probability of obtaining the outcome $ k $ is given by the Born rule, i.e., $p(k|P, M)=Tr[\rho E_{k}]$. 

The ontological model of quantum theory can be described as follows \cite{hari,spek05}. Whenever the density matrix $\rho$ is prepared by the preparation procedure $P$, it is assumed that in an ontological model a probability distribution $\mu_{P}(\lambda|\rho)$ of the ontic state $\lambda\in \Lambda$ is prepared. Here, $\Lambda$ is the ontic state space. The probability distribution is normalized, i.e.,   $\int _\Lambda \mu_{P}(\lambda|\rho)d\lambda=1$. Now, when the measurement of $E_{k}$ is performed through a measurement procedure $M$,  the ontic state  assigns the probability of obtaining the outcome $k$ known as response function $\xi_{M}(k|\lambda, E_{k}) $, satisfying $\sum_{k}\xi_{M}(k|\lambda , E_{k})=1$. A viable ontological model must reproduce the quantum probabilities. Hence, $\forall \rho $, $\forall E_{k}$ and $\forall k$, $\int _\Lambda \mu_{P}(\lambda|\rho) \xi_{M}(k|\lambda, E_{k}) d\lambda =Tr[\rho E_{k}]$.

 Now, the notion of noncontextuality in an ontological model arises from the equivalence class of experimental procedures. As argued in \cite{spek05} that operationally equivalent experimental procedures can be equivalently represented in an ontological model.  For example, if two preparation procedures $P$ and $P^{\prime}$ prepare the same density matrix $\rho$, then no measurement can operationally distinguish the context by which $\rho$ is prepared.  This leads to the notion of preparation noncontextuality in an ontological model of quantum theory, i.e.,  
\begin{align}
\label{ass}
\forall M, \  k: \ \   p(k|P, M)=p(k|P^{\prime},M)	\Rightarrow \mu_{P}(\lambda|\rho)=\mu_{P^{\prime}}(\lambda|\rho)
\end{align}
 which implies that the distribution of ontic states are equivalent irrespective of the contexts $ P $ and $ P^{\prime}$ \cite{spek05,pan19}. Below we argue that in a preparation noncontextual ontological model the parity-oblivious constraint in a communication game implies an equivalent obliviousness condition at the level of ontic states. 
 


\subsection{Parity-oblivious communication game}

Let in an operational theory, Alice prepares the inputs $x$ by the preparation procedures $P_{x}$, and upon receiving the input $y$, Bob performs the measurement of $M_{y}$. Consider that there are $L$ subsets having equal number of elements of the input $P_{l}\subset P_{x}$ with $l=1,2,3 . . ..L$. Let us now impose a constraint in the preparation procedures in terms of obliviousness condition.  Such a restriction  demands that given an input is not distinguishable by any measurement whether it has come from $P_{l}\subset P_{x}$ or from $P_{l^{\prime}}\subset P_{x}$. This holds even when  Alice's amount of communication is not restricted. For our purpose, it will be enough to consider the inputs of Alice are uniformly distributed, so that $p_{A}(x)=1/|P_{x}|$ where $|P_{x}|$ is the cardinality of the set. Then,  $	\forall \ l, l^{\prime}, y, b $, the oblivious condition implies
\begin{align}
\label{po11}
 \sum\limits_{P_{x}\in P_{l}} p(P_{x}|b,M_{y}) = \sum\limits_{P_{x}\in P_{l^{\prime}}} p(P_{x}|b,M_{y}) 
\end{align}
Using the Bayes rule $p(P_x|b,M_y)=p(b|P_x,M_y) p(P_x|M_y)/p(b|M_y)$,
  Eq. (\ref{po11}) can be written as  
 
\begin{align}
\label{po111}
\forall b, y: \ \ \sum\limits_{P_{x}\in P_{l}} p(b|P_{x},M_{y}) = \sum\limits_{P_{x}\in P_{l^{\prime}}} p(b|P_{x},M_{y}) 
\end{align}This then implies that in an operational theory the two preparation procedures $P_{l}$ and $P_{l^{\prime}}$  cannot be distinguishable for any outcome $b$ and the measurement $M_{y}$ of Bob.  As argued by Spekkens \cite{spek05}, that the operationally equivalent experimental procedures can be equivalently represented in an ontological model of the said operational theory. Such an equivalence introduces the notion of noncontextuality in an ontological model.

 Hence, for an ontological model of the operational theory satisfying the assumption of preparation non-conextuality \cite{spek05,pan19,ravi,hameedi17b} we can write
\begin{align}
\label{pnc}
 \sum\limits_{{P_{x}}\in P_{l}} \mu(\lambda|P_{x}) =  \sum\limits_{{P_{x}}\in P_{l^{\prime}}} \mu(\lambda|P_{x})
\end{align}
where $\lambda \in \Lambda$ is the ontic state and $\Lambda$ is the ontic state space.
Using Bayes theorem and noting the uniform distributon of $P_{x}$, we have
\begin{align}
\label{}
\sum\limits_{P_{x}\in P_{l}} \mu(P_{x}|\lambda) =  \sum\limits_{P_{x}\in P_{l^{\prime}}} \mu(P_{x}|\lambda)
\end{align}
This implies that the obliviousness condition must also be satisfied at the level of ontic states $\lambda$ for the preparation noncontextual model. In this work, we specifically focus on the PORAC in which no parity information of the inputs will be transmitted to Bob due to Alice's communication.
\subsection{$n$-bit two-level PORAC}

In a standard $2$-level $n$-bit RAC, Alice receives length-$n$ strings  randomly sampled from $x\in \{0,1\}^n$. Bob receives, uniformly at random, the numbers $y \in [n]$. Bob's task is to recover $y^{\text{th}}$ bit of Alice's input string $x$, which is the winning condition of the game. Throughout this paper, we restrict ourselves in a $2$-level RAC. 
	
In an arbitrary operational theory, Alice encodes her input string $x$ in a physical state by preparation procedure $P_{x}$, and send it to Bob.  After receiving the state Bob performs a binary outcome measurement $M_{y}$ for every   $y=\{1,2,...,n\}$ and reports outcome $b$ as his output. Given the winning condition $b=x_{y}$, the success probability of the RAC is given by
	\begin{align}
		\label{SP}
		\mathcal{P}_n=\dfrac{1}{2^{n}n}\sum_{y=1}^{n}\sum_{x\in \{0,1\}^{n}}p(b=x_{y}|P_{x},M_{y}).
	\end{align}
	
		Since its a collaborative game, Alice communicates some bits of information to Bob to help him. However, in PORAC we impose the parity-oblivious restriction on Alice's inputs so that her communication to Bob must not reveal the parity information of her inputs. In \cite{spek09}  the parity-oblivious condition is defined corresponding to a particular parity-set as
	\begin{align}
	\label{gn2}
		\mathbb{G}_{n,2}=\{x|x \in \{0,1\}^{n} , \sum_{r=1}^{n}x_{r} \geq 2\}
	\end{align}
	 The parity-oblivious condition demands that for every element $s \in \mathbb{G}_{n,2}$, no information about $s.x=\oplus_{r}s_{r}x_{r}$ is to be transmitted to Bob ($\oplus_{r}$ denotes sum modulo $2$), and every element $s$ introduces a constraint on the inputs.  For every $s$, the input strings will be divided to $s$-parity 0 and $s$-parity 1 sets corresponding to $s.x=0$ and  $s.x=1$ respectively. From Alice's communication, Bob should not be able to determine  in which set a specific bit string belongs. Average success probability in classical PORAC is \cite{spek09} 
			\begin{align}
			\label{pncb}
				(\mathcal{P}_{n})_{cl}\leq \frac{(n+1)}{2n}.
			\end{align}
	The detailed derivation of this bound can be found in \cite{spek09}. We note here that the way the parity-set  in Eq. (\ref{gn2}) is defined, \emph{at most} one bit can be communicated. Else, it will reveal the parity information.  Intuitively, if Alice agrees beforehand to reveal the first bit ($x_{1}$), then for $y=1$, occurring with probability $1/n$, Bob can predict $b=x_{y}$ with certainty. For $y\neq 1$, occurring with probability $(n-1)/n$, Bob at best guesses the bit $x_{y}$ with probability $1/2$. Then average success probability is $1/n +(n-1)/2n$, saturating the above bound in Eq. (\ref{pncb}). 
	
\subsection{$n$-bit two-level quantum PORAC}

	In a quantum RAC, Alice encodes her input string $x$ into density matrix $\rho_{x}$, prepared by a procedure $P_{x}$. Bob performs a binary measurement $\{E_{y}^{b}\}$ (with $\sum_{b}E_{y}^{b}=\mathbb{I}$ and $b=0,1$) for every bit of $x$ he wants to decode. The quantum success probability can then be written as 
	\begin{equation}
		\label{QSP}
		(\mathcal{P}_n)_{Q}=\dfrac{1}{2^{n}n}\sum_{y=1}^{n}\sum_{x\in \{0,1\}^{n}}\mathrm{tr}\hspace{2pt}[\rho_{x}E_{y}^{b}].
	\end{equation}
	Here $\rho_{x}$ are the density matrices prepared by procedure $P_{x}$ by performing local measurement on Alice's set of observables.  
	
Analogously, in quantum PORAC the parity-oblivious condition demands 			\begin{align}
				\label {po3}
				\forall s: \  \dfrac{1}{2^{n-1}}\sum\limits_{x|x.s=0}\rho_{x} =\dfrac{1}{2^{n-1}}\sum\limits_{x|x.s=1}\rho_{x}
			\end{align}
		to be satisfied. This is then within the premise of the preparation noncontextuality in an ontological model \cite{spek05} of operational quantum theory.  Assuming preparation noncontextuality in an ontological model, the parity-oblivious condition can be equivalently represented in an ontological model, so that 
\begin{align}
\forall s: \  \dfrac{1}{2^{n-1}}\sum\limits_{x|x.s=0}\mu(\lambda| \rho_{x}) =\dfrac{1}{2^{n-1}}\sum\limits_{x|x.s=1}\mu(\lambda| \rho_{x}), 
\end{align}
where $\lambda \in \Lambda$ is the ontic state and $\Lambda$ is the ontic state space. Following earlier discussions, by using Bayes rule it can be shown that 
\begin{align}
\forall s: \  \dfrac{1}{2^{n-1}}\sum\limits_{x|x.s=0}\mu( \rho_{x}|\lambda) =\dfrac{1}{2^{n-1}}\sum\limits_{x|x.s=1}\mu( \rho_{x}|\lambda), 
\end{align} 
implying that for preparation noncontextual models, the satisfaction of parity-obliviousness condition in an operational theory provides equivalent representation at the level of the ontic state $\lambda$ and it cannot contain any information about the parity. Owing to the above discussion, Spekkens \emph{et. al.} \cite{spek09} argued that the bound on the average success probability for a PORAC given in Eq. (\ref{pncb}) can also be said preparation noncontextual bound $(\mathcal{P}_{n})_{pnc}$. Moreover, the optimal quantum success probability of a $n$-bit standard PORAC from Eq. (\ref{QSP}) is derived as $(\mathcal{P}_n)_Q^{opt}=(1/2)(1+1/\sqrt{n})> (\mathcal{P}_n)_{pnc}$ \cite{ghorai}. Interestingly, the success probability decreases monotonically with increasing $n$. 

In this work, we propose a nontrivial generalization of the $n$-bit PORAC by  introducing a generalized parity-set. We derive the optimal success probability of that generalized PORAC in a preparation noncontextual ontological model for any arbitrary $n$. Further, we demonstrate the quantum supremacy  over preparation noncontextual model in a prepare-and-measure scenario of the communication game.

\section{Generalized PORAC and preparation noncontextual bound} We note again that in Spekkens \emph{et al.} \cite{spek09}, the PO condition (in Eq. (\ref{gn2})) imposed on Alice's inputs is based on a particular parity-set where each element $s \in \mathbb{G}_{n,2} $ gives a different constraint on the inputs.   We propose a nontrivial generalization of the standard PORAC \cite{spek09} by introducing the parity-set  
\begin{align}
\label{gpo}
\mathbb{G}_{n, g_n}=\{x|x\in \{0,1\}^{n} , \sum_{r=1}^{n}x_{r} \geq g_n\}
\end{align}
where $g_n \in \{2,3..., n \}$. The PO condition demands that for every element $s \in \mathbb{G}_{n,g_n}$, no information about $s.x=\oplus_{r}s_{r}x_{r}$ is to be transmitted to Bob ($\oplus_{r}$ denotes sum modulo $2$), and every element $s$ introduces a constraint on the inputs.  For every $s$, the input strings will be divided to $s$-parity 0 and $s$-parity 1 sets corresponding to $s.x=0$ and  $s.x=1$ respectively. For $g_n=2$, one recovers Eq. (\ref{gn2}) the standard PORAC in Spekkens \emph{et. al.} \cite{spek09}. 

In a $n$-bit PORAC, the PO condition demands that the classical message $m$ sent from Alice to Bob must satisfy
\begin{equation}
\label{eqq1}
\forall s \in  \mathbb{G}_{n, g_n} \ : \sum_{x|x.s=0}p(P_{x}|m)=\sum_{x|x.s=1}p(P_{x}|m).\end{equation} By noting the fact that in a preparation noncontextual ontological model, the PO condition is equivalently represented, We propose the following theorem.\\
			
{\emph{\textbf{Theorem 1:} For any arbitrary $n$ and $g_n$ the optimal success probability in a preparation noncontextual model is 
\begin{align}
\label{gpnc}
(\mathcal{P}_{n}^{g_n})_{pnc}=\frac{(g_n+n-1)}{2n}.
\end{align}	}			
The proof of the above Theorem 1 is quite technical. We provide detailed proofs for $n=3$ in Appendix \ref{app3}, and for arbitrary $n$ in Appendix \ref{proofmn}. We also provide a simple strategy that reproduces the optimal value in Eq. (\ref{gpnc}) for arbitrary $n$.

An intuitive argument for the simplest case of $3$-bit game could be useful here. As defined in Eq. (\ref{gpnc}), for $g_3=3$ the relevant parity-set $\mathbb{G}_{3, 3}$ contains only one element  $s=111$. Then, the  $s$-parity $0$ and $s$-parity $1$ sets are $\{000, 011, 101,110\}$ and $\{001, 010, 100, 111\}$. It is proved in Appendix \ref{app3} that Alice is allowed to communicate at most two bits to Bob. Considering the pre-decided strategy, Alice communicates the first two bits of her input. It can be checked that this will not reveal the parity information of her input. Then for both the first and second bits appearing with the probability $1/3$, Bob is certain about the bit, but for the third bit, he guesses randomly, i.e., with probability $1/2$.  Thus, the success probability in a preparation noncontextual bound is derived as $(\mathcal{P}_{3}^{g_3=3})_{pnc}=5/6$. A similar argument holds for any arbitrary two bits. It remains to explicitly prove that Alice's communication of two bits of information is required to obtain the optimal preparation noncontextual value which we have provided in Appendix \ref{app3} along with the preparation noncontextual bound for $n=3$.


We argue that for any arbitrary $n$ and $g_n$, Alice can communicate  $(g_n-1)$ bits without revealing the parity information. Then, for $y=\{1,2,...,g_n-1\}$, each occurring with probability $1/n$, Bob can predict the outcome $b=x_{y}$ with certainty, and for rest of the bits $(n-g_n+1)$, Bob can only randomly guess the bit with probability $1/2$. Thus, the average success probability in the preparation noncontextual becomes is $(g_n-1)/n + (n-g_n+1)/2$, as claimed in Eq. (\ref{gpnc}). For $g_n=2$, the preparation noncontextual bound in \cite{spek09} is recovered. 

We show that for any arbitrary $n$-bit PORAC corresponding to the parity-set $\mathbb{G}_{n, g_n}$, quantum theory outperforms the classical noncontextual model. Given the $g_n$, in a quantum PORAC Alice encodes her $n$-bit strings into density matrices $\rho_{x} \in 2^{\otimes(g_n-1)}$.  We shortly show that  $\rho_{x}$s have to be pure states to get an optimal quantum advantage.\\

{\emph {\textbf{Results:} The optimal quantum success probability violates the preparation noncontextual bound of the generalized PORAC corresponding to the parity-set $\mathbb{G}_{n, g_n}$ for any arbitrary $n$, i.e., $ (\mathcal{P}_{n}^{g_n})_{Q}>(\mathcal{P}_{n}^{g_n})_{pnc}$} .
		
Before proving the general results for any arbitrary $n$ and $g_n$, for the sake of better understanding,  we explicitly derive the quantum advantage of our generalized PORAC for $g_3=3$ and $g_4=3,4$.  Further, we demonstrate the quantum advantage for the generalized quantum PORAC for $g_n=n$ when $n>4$. Note that, those derivations are fully analytical, but for the higher values of $n$ for arbitrary $g_{n}$ one requires computational work.

\section{$3$-bit generalized quantum PORAC}	

For $3$-bit PORAC, we have two games corresponding to the two parity sets, $\mathbb{G}_{3, 2}$ (the standard one) and $\mathbb{G}_{3, 3}$ for $g_{3}=2$ and $3$ respectively. The case of $\mathbb{G}_{3, 2}$ is already considered in \cite{spek09}. We explicitly demonstrate the PORAC corresponding to $\mathbb{G}_{3, 3}$ where the preparation noncontextual bound is  $(\mathcal{P}_{3}^{g_{3}=3})_{pnc}=5/6$. Also,  Alice is allowed to communicate at most two bits of information to Bob to satisfy the PO condition, i.e., without revealing the parity information of her input.

Note that, for the parity-set $\mathbb{G}_{3,3}$ in Eq. (\ref{gpnc}), we have only one parity element  $s=111$, leading to the  $s$-parity $0$ set is $\{000, 011, 101,110\}$ and the $s$-parity $1$ set is $\{001, 010, 100, 111\}$. The PO condition in an operational theory needs to satisfy the following condition. 
\begin{eqnarray} \label{po}\nonumber
&&p(P_{000}|b,M_{y})+p(P_{110}|b,M_{y})+p(P_{011}|b,M_{y})\\ \nonumber
&+&p(P_{101}|b,M_{y})=p(P_{100}|b,M_{y})+p(P_{010}|b,M_{y})\\ 
&+&p(P_{001}|b,M_{y})+p(P_{111}|b,M_{y})
\end{eqnarray}

In quantum theory, Alice encodes her inputs $x\in \{0,1\}^{3}$ into eight quantum states $\rho_{x}\in \mathbb{C}^{2} \otimes \mathbb{C}^{2}$ and sends them to Bob.  The PO condition corresponding to the parity element  $s=111$ then provides 
 \begin{eqnarray} \label{poq}
\rho_{000}+\rho_{011}+\rho_{110}+\rho_{101}=\rho_{001}+\rho_{010}+\rho_{100}+\rho_{111}
 \end{eqnarray} i.e., the mixture of the four states corresponding to the even parity is identical to the mixture of the four states corresponding to odd parity. After receiving the particles, Bob performs three projective measurements $ \{\Pi_{3,y}^{0},\Pi_{3,y}^{1}\}$ with $y=1,2,3$, with $b= x_y\in \{0,1\}$. The quantum success probability from Eq. (\ref{QSP}) can be explicitly written as
\begin{widetext}
\begin{eqnarray}
	\label{A1m}
	\nonumber
	(\mathcal{P}_{3})_{Q}=  \dfrac{1}{24}&\mathrm{tr}&\Big[p(0|\rho_{000},\Pi_{3,1}^{0})+p(0|\rho_{000},\Pi_{3,2}^{0})+p(0|\rho_{000},\Pi_{3,3}^{0})+ p(1|\rho_{111},\Pi_{3,1}^{1})+p(1|\rho_{111},\Pi_{3,2}^{1})+p(1|\rho_{111},\Pi_{3,3}^{1})\\
	&+&p(0|\rho_{010},\Pi_{3,1}^{0})+p(1|\rho_{010},\Pi_{3,2}^{1})+p(0|\rho_{010},\Pi_{3,3}^{0}) +p(1|\rho_{101},\Pi_{3,1}^{1})+p(0|\rho_{101}\Pi_{3,2}^{0})+p(1|\rho_{101},\Pi_{3,3}^{1})\\
	\nonumber
	&+&p(0|\rho_{001},\Pi_{3,1}^{0})+p(0|\rho_{001},\Pi_{3,2}^{0})+p(1|\rho_{001},\Pi_{3,3}^{1})+ p(1|\rho_{110},\Pi_{3,1}^{1})+p(1|\rho_{110},\Pi_{3,2}^{1})+p(0|\rho_{110},\Pi_{3,3}^{0})\\
	\nonumber
	&+&p(0|\rho_{011},\Pi_{3,1}^{0})+p(1|\rho_{011},\Pi_{3,2}^{1})+p(1|\rho_{011},\Pi_{3,3}^{1})+p(1|\rho_{100},\Pi_{3,1}^{1})+p(0|\rho_{100},\Pi_{3,2}^{0})+p(0|\rho_{100},\Pi_{3,3}^{0})\Big]
\end{eqnarray} 
\end{widetext}
Using  $\Pi_{3,y}^{0} + \Pi_{3,y}^{1} = \mathbb{I}$ with $y=1,2,3$, Eq. (\ref{A1m}) can be re-written as
\begin{eqnarray}
	\label{eqn9}
	\nonumber
	(\mathcal{P}_{3})_{Q}= \dfrac{1}{2}+\dfrac{1}{24}&\mathrm{tr}&\Big[(\rho_{000}-\rho_{111})(\Pi_{3,1}^{0}+\Pi_{3,2}^{0}+\Pi_{3,3}^{0})\\
	\nonumber&+&(\rho_{010}-\rho_{101})(\Pi_{3,1}^{0}-\Pi_{3,2}^{0}+\Pi_{3,3}^{0})\\
	\nonumber
	&+&(\rho_{001}-\rho_{110})(\Pi_{3,1}^{0}+\Pi_{3,2}^{0}-\Pi_{3,3}^{0})\\
	\nonumber
	&+&(\rho_{011}-\rho_{100})(\Pi_{3,1}^{0}-\Pi_{3,2}^{0}-\Pi_{3,3}^{0})\Big]\\
\end{eqnarray}
which can be further simplified by using $\Pi_{3,y}^{b} = (\mathbb{I}+(-1)^{b}B_{3,y})/{2}$ where $b\in \{0,1\}$ and by suitably rearranging as
\begin{eqnarray}
\label{eqnA5m}
\nonumber
(\mathcal{P}_{3})_{Q}=\dfrac{1}{2}&+&\dfrac{1}{48}\mathrm{tr}\Big[(\rho_{000}-\rho_{110}+\rho_{011}-\rho_{101})B_{3,1}\\
 \nonumber
 &+&(\rho_{000}-\rho_{110}-\rho_{011}+\rho_{101})B_{3,2}\\
  \nonumber
 &+&(\rho_{000}+\rho_{110}-\rho_{011}-\rho_{101})B_{3,3}\\
 \nonumber
 &+&(\rho_{001}+\rho_{010}-\rho_{111}-\rho_{100})B_{3,1}\\
 &+&(\rho_{001}-\rho_{010}-\rho_{111}+\rho_{100})B_{3,2}\\
 \nonumber
 &-&(\rho_{001}-\rho_{010}+\rho_{111}-\rho_{100})B_{3,3}\Big].
\end{eqnarray}

To obtain optimal quantum success probability, let us first analyze the term $\mathrm{tr}[(\rho_{000}-\rho_{110}+\rho_{011}-\rho_{101}) B_{3,1}]$. Note that to get maximum value  
$\rho_{000}$ and  $\rho_{011}$ have to be the eigenstates of $B_{1}$ with eigenvalue $+1$, and $\rho_{110}$ and $\rho_{101}$ have to be  eigenstates of $B_{1}$ with eigenvalue $-1$. Since we consider two-qubit system, $\rho_{000}, \rho_{011}, \rho_{110}$ and $\rho_{101}$  has to be orthogonal pure states implying $\rho_{000}+\rho_{110}+\rho_{011}+\rho_{101}=\mathrm{I}\otimes \mathrm{I}$. Similar argument can be made for the second and third terms. Now, if we consider $\mathrm{tr}[(\rho_{001}+\rho_{010}-\rho_{111}-\rho_{100})B_{3,2}]$, maximization criteria gives $\rho_{001}, \rho_{010},\rho_{111}$ and $\rho_{100}$ to beorthogonal pure states, ie., $\rho_{001}+\rho_{010}+\rho_{111}+\rho_{100}=\mathrm{I}\otimes \mathrm{I}$ . This in turn satisfies the PO condition in Eq. (\ref{poq}). The same argument holds for last two terms.

Above argument leads us to define three two-qubit observables of the form 
\ba 
\nonumber
A_{3,1} = \rho_{000}+\rho_{011}-\rho_{110}-\rho_{101}\\
A^{\prime}_{3,1} = \rho_{000}-\rho_{011}+\rho_{110}-\rho_{101}\\
\nonumber
\tilde{A}_{3,1}=\rho_{000}-\rho_{011}+\rho_{110}-\rho_{101}
\ea
By construction  $[A_{3,1},A^{\prime}_{3,1}]=[A^{\prime}_{3,1},\tilde{A}_{3,1}]=[A_{3,1},\tilde{A}_{3,1}]=0$, as $A_{3,1}, A^{\prime}_{3,1}$ and $\tilde{A}_{3,1}$ have common eigenstates. 

Similarly, we can construct three more two-qubit observables corresponding to the other four states $\rho_{x}$ for which $x|x.s=1$, are given by
\ba A_{3,2}= \rho_{001}+\rho_{010}-\rho_{100}-\rho_{111}\\
\nonumber
A^{\prime}_{3,2}=\rho_{001}-\rho_{010}+\rho_{100}-\rho_{111}\\
\nonumber
\tilde{A}_{3,2}=\rho_{001}-\rho_{010}-\rho_{100}+\rho_{111}
\ea
where the observables satisfies $[A_{3,2},A^{\prime}_{3,2}]=[A^{\prime}_{3,2},\tilde{A}_{3,2}]=[A_{3,2},\tilde{A}_{3,2}]=0$.

Note here that, since in two-qubit system there exists at most three mutually commuting observables, the following relations must be satisfied
\ba
\label{constraint}
A_{3,1}A^{\prime}_{3,1}=\pm {\tilde A}_{3,1}; \nonumber\\ A_{3,2}A^{\prime}_{3,2}=\pm {\tilde A}_{3,2}. 
\ea
This is due to the fact that  the product of any two observables has to return the third one with $\pm$.


Thus, the quantum success probability from Eq. (\ref{eqnA5m}) can be written as 
\begin{equation}
\label{sp3-3}
(\mathcal{P}_{3}^{g_3=3})_Q=\dfrac{1}{2}+\dfrac{\mathrm{tr}[\Delta_3^{g_3=3}]}{48}
\end{equation}
where $\Delta_{3}^{g_3=3}$ is the correlation function of the form
\begin{eqnarray}
\label{delta33}
\Delta_{3}^{g_3=3}&=&(A_{3,1}+A_{3,2}) B_{3,1}\\
\nonumber
&+& ({\tilde A}_{3,1}-{\tilde A}_{3,2})B_{3,3}+ (A^{\prime}_{3,1}+A^{\prime}_{3,2})B_{3,2}. 
\end{eqnarray}

For deriving the quantum bound, without loss of generality we can impose one of the three conditions $A_{3,1}=A_{3,2}$, $A^{\prime}_{3,1}=A^{\prime}_{3,2}$ and ${\tilde A}_{3,1}=-{\tilde A}_{3,2}$. This is due to the fact that if we choose, say, $A^{\prime}_{3,1}=A^{\prime}_{3,2}\equiv A^{\prime}_{3}$ so that both sets $\{\rho_{x}|_{s.x=1}\}$ and $\{\rho_{x}|_{s.x=0}\}$ are the eigenstates of $A^{\prime}_{3}$. It can be simply checked that this does not affect the preparation noncontextual bound.  Such choice of observables are available in two-qubit system. 

Optimal value is achieved when  $B_{3,1} = (A_{3,1} +A_{3,2})/\sqrt{2}$ , $B_{3,2} = A^{\prime}_{3}$ and $B_{3,3} = (\tilde{A}_{3,1} -{\tilde A}_{3, 2})/\sqrt{2}$. Thus, the optimal quantum value of the correlation function $\Delta_{3}^{g_3=3}$ in Eq. (\ref{delta33}) is derived as 
\ba \mathrm{tr}[\Delta_{3}^{g_3=3}]_{Q}^{opt} = 8+8\sqrt{2}.\ea 
which in turn provides the optimal quantum success probability 
\begin{equation}	(\mathcal{P}_{3}^{g_3=3})_{Q}^{opt}=\dfrac{1}{2}+\dfrac{1+\sqrt{2}}{6}\approx0.902
\end{equation} 
i.e., outperforming the preparation noncontextual bound $5/6 \approx 0.833$. 

One of such the choices of Alice's  observables are 
\begin{eqnarray}
\nonumber
    A_{3,1} = \sigma_{x} \otimes \sigma_{x}&&,\  \ \ \  \tilde{A}_{3,1} = \mathbb{I} \otimes \sigma_{x};\ \ A^{\prime}_{3} = \sigma_{x} \otimes \mathbb{I};\\
 A_{3,2}= \sigma_{x} \otimes \sigma_{z}&&,\  \ \ \  \tilde{A}_{3,2} = \mathbb{I} \otimes \sigma_{z},
\end{eqnarray}
satisfying the required mutual commutation relation of the sets  $\{A_{3,1},A^{\prime}_{3}, {\tilde A}_{3,1}\}$ and  $\{A_{3, 2},A^{\prime}_{3}, {\tilde A}_{3,2}\}$. Hence, the encoding density matrices can be easily constructed from the commuting observables as $\rho_{000}=|+0\rangle \langle 0+|$, $\rho_{110}=|+1\rangle \langle 1+|$, $\rho_{001}=|++\rangle \langle ++| $, $\rho_{100}=|-+\rangle \langle +-|$ and so on.

Note that maximum quantum value of the sequential correlation, say, $\mathrm{tr}[A_{3,1} B_{3,1}]$ is $4$, as $A_{3,1}$ and $B_{3,1}$ are two-qubit observables. By keeping this in mind while deriving the bound in preparation noncontextual model, one may surmise that $\mathrm{tr}[\Delta_{3}^{g_3=3}]$ has the classical upper bound $24$. But, in a preparation noncontextual model, $\mathrm{tr}[\Delta_{3}^{g_3=3}]_{pnc}\leq 16$. Consequently, we have $\left(\mathcal{P}_3^{g_3=3}\right)_{pnc}\leq 5/6\approx0.833$, as claimed in Eq. (\ref{gpnc}). This is due to the fact that to satisfy the PO condition, the relations in Eq. (\ref{constraint}) has to be satisfied.  Owing to this notion of equivalent representation of quantum theory in a preparation noncontextual ontological model, $\xi(k|A_{3,1},\lambda)\xi(k|A^{\prime}_{3,1},\lambda)=\pm \xi(k|{\tilde A}_{3,1},\lambda)$ and $\xi(k|A_{3,2},\lambda)\xi(k|A^{\prime}_{3,2},\lambda)=\pm \xi(k|{\tilde A}_{3,2},\lambda)$. This restriction ensures $\mathrm{tr}[\Delta_{3}^{g_3=3}]_{pnc}\leq 16$.

In order to test our proposed $3$-bit PORAC game in the prepare-and-measure scenario, we provide an experimental setup. We consider joint spin (polarization) and path degrees of freedom of a single particle and suitable Mach-Zehender interferometer along with phase shifters and beam splitters for implementing our protocol. The details of the setup with necessary illustration are placed in  Appendix \ref{experimet}. 
\subsection{$3$-bit generalized PORAC in entanglement-assisted scenario}
Our proposed prepare-and-measure scenario can also be played in an entanglement-assisted scenario as shown in Appendix \ref{app3ea}. Note that satisfying the PO condition,  Alice can at most communicate two bits of classical information to Bob. In Appendix \ref{app3ea}, we provide a detailed encoding scheme using sequential degeneracy-breaking measurement of two commuting observables as well as the decoding scheme where Bob performs projective measurement for the optimal quantum success probability in $3$-bit PORAC game. 

We demonstrate that the optimal quantum success probability remains the same as in the prepare-and-measure scenario. We note here that, in the entanglement-assisted scenario, although we consider two-qubit system to demonstrate quantum supremacy, the quantum success probability of $3$-bit PORAC game can be derived without considering the dimension of the system. This feature enables a device-independent self-testing based on our generalized PORAC game. We also provide an encoding-decoding strategy that can reestablish the optimal quantum success probability in terms of the optimal quantum violation of a suitably derived Bell expression $\mathcal{B}_{3}^{g_3=3}$ in Appendix \ref{app3ea}. 

\section{$4$-bit generalized quantum PORAC} 
\label{4bitgen}

Before generalizing for arbitrary $n$ and $g_n$, we consider $4$-bit PORAC where we have three different games corresponding to the parity sets $\mathbb{G}_{4,2}$ (standard one), $\mathbb{G}_{4,3}$ and $\mathbb{G}_{4,4}$. The game corresponding to $\mathbb{G}_{4,2}$ is the standard one which has  already been explored in \cite{spek09,ghorai}.

\subsection{Optimal quantum advantage for $\mathbb{G}_{4,4}$}
We start by noting that for $\mathbb{G}_{4,4}$, the parity-set contains only one element $s=1111$. In such a case, Alice is allowed to communicate three bits of information to Bob without revealing the parity information of her input. The preparation noncontextual bound is already calculated as $(\mathcal{P}_{4}^{g_4=4})_{pnc}=7/8$. 

In quantum theory, Alice encodes the inputs in sixteen quantum states $\rho_{x} \in \mathbb{C}^{2} \otimes \mathbb{C}^{2} \otimes \mathbb{C}^{2}$. The PO condition corresponding to the parity element $s=1111$ requires $\sum_{x|s.x=0}\rho_{x}=\sum_{x|s.x=1}\rho_{x}$ to be satisfied. Now, employing a similar strategy as in the case of $3$-bit PORAC game we obtain the optimal quantum success probability. The detailed derivation is provided in Appendix \ref{optm44}.

We again argue that the maximum value requires $\rho_{x}$'s with $s.x=0$ to be orthogonal eigenstates of $B_{4,1}$ which in turn satisfies $\sum_{x|s.x=0}\rho_{x}=\mathrm{I}\otimes\mathrm{I}\otimes\mathrm{I}$. This leads us to define four three-qubit observables having common eigenstates $A_{4,1},A^{\prime}_{4,1},A^{\prime\prime}_{4,1}$ and $\tilde{A}_{4,1}$ satisfying $[A_{4,1},A^{\prime}_{4,1}]=[A^{\prime}_{4,1},A^{\prime\prime}_{4,1}]=[A^{\prime\prime}_{4,1},\tilde{A}_{4,1}]=[A_{4,1},\tilde{A}_{4,1}]=0$ and so on. Similar argument holds good for $\rho_{x}$ with $s.x=1$ where $\sum_{x|s.x=1}\rho_{x}=\mathrm{I}\otimes\mathrm{I}\otimes\mathrm{I}$ will satisfy such that four three-qubit observables having common eigenstates $A_{4,2},A^{\prime}_{4,2},A^{\prime\prime}_{4,2}$ and $\tilde{A}_{4,2}$ can be defined.
  
  The decoding scheme is also similar to $3$-bit case. As derived in Appendix \ref{optm44}, the quantum success probability is given by
\begin{equation}
\label{sp4-4}
(\mathcal{P}_{4}^{g_4=4})_Q=\dfrac{1}{2}+\dfrac{\mathrm{tr}[\Delta_4^{g_4=4}]_Q}{128}
\end{equation}
			where the expression $\Delta_4^{g_4=4}$ is derived as 
			\begin{eqnarray}\label{del44}
			\nonumber
				\Delta_4^{g_4=4}&=&(A_{4,1}+A_{4,2}) B_{4,1}+( \tilde{A}_{4,1}-\tilde{A}_{4,2}) B_{4,4}\\
					\label{delta4}
				&+& (A^{\prime}_{4,1}+A^{\prime}_{4,2}) B_{4,2}+ (A^{\prime\prime}_{4,1}+A^{\prime\prime}_{4,2}) B_{4,3}.
			\end{eqnarray}

Here $\tilde{A}_{4,1}$ is commuting with $A_{4,1}, A^{\prime}_{4,1}$ and $A^{\prime\prime}_{4,1}$ and hence $A_{4,1}A^{\prime}_{4,1}A^{\prime\prime}_{4,1}=\pm\tilde{A}_{4,1}$. This comes from the fact that in a three-qubit system there are at most four mutually commuting observables. Similarly, we have $A_{4,2}A^{\prime}_{4,2}A^{\prime\prime}_{4,2}=\pm\tilde{A}_{4,2}$. In an ontological model of quantum theory this implies $\xi(k|A_{4,1},\lambda)\xi(k|A^{\prime}_{4,1},\lambda)\xi(k|A^{\prime\prime}_{4,1},\lambda)=\pm \xi(k|{\tilde A}_{4,1},\lambda)$ and $\xi(k|A_{4,2},\lambda)\xi(k|A^{\prime}_{4,2},\lambda)\xi(k|A^{\prime\prime}_{4,2},\lambda)=\pm \xi(k|{\tilde A}_{4,2},\lambda)$.  Using them in Eq. (\ref{del44}),  we derive $\mathrm{tr}[\Delta_{4}^{g_4=4}]_{pnc}\leq 48$ and consequently from Eq. (\ref{sp4-4}) $(\mathcal{P}_4^{g_4=4})_{pnc}=7/8\approx 0.875$.  This eventually certifies the bound calculated for $g_{4}=4$ in Eq. (\ref{gpnc}).
   
Now, by keeping in mind that we are dealing with a three-qubit system  and with the aim of optimization of the correlation function Eq. (\ref{del44}), without loss of generality, we can always choose $A^{\prime}_{4,1}=A^{\prime}_{4,2}=A^{\prime}_{4}$ and $A^{\prime\prime}_{4,1}=A^{\prime\prime}_{4,2}=A^{\prime\prime}_{4}$ for deriving the quantum optimal bound. This is due to the fact that if we choose, say, $A^{\prime}_{4,1}=A^{\prime}_{4,2}\equiv A^{\prime}_{4}$ so that both sets $\{\rho_{x}|_{s.x=1}\}$ and $\{\rho_{x}|_{s.x=0}\}$ are the eigenstates of $A^{\prime}_{4}$. Similarly, we can choose $A^{\prime\prime}_{4,1}=A^{\prime\prime}_{4,2}\equiv A^{\prime\prime}_{4}$. This means we have two sets of commuting observables $\{A_{4,1}, \tilde{A}_{4,1},A^{\prime}_{4},A^{\prime\prime}_{4}\}$ and $\{A_{4,2}, \tilde{A}_{4,2},A^{\prime}_{4},A^{\prime\prime}_{4}\}$. It can be simply checked that the preparation noncontextual bound remains unchanged.  

   With the above choices of observables, we can rewrite the correlation function in Eq. (\ref{del44}) as

   \begin{eqnarray}\label{del441}
\nonumber
\Delta_4^{g_4=4}=&&(A_{4,1}+A_{4,2}) B_{4,1}+( \tilde{A}_{4,1}-\tilde{A}_{4,2}) B_{4,4}\\
\label{delta4}
&+& 2A^{\prime}_{4} B_{4,2}+ 2 A^{\prime\prime}_{4} B_{4,3}.
			\end{eqnarray}

 whose optimal quantum value is derived as \begin{equation}
			    \mathrm{tr}[\Delta_4^{g_{4}=4}]_Q^{opt}= 32+16\sqrt{2}
			\end{equation}
   This is obtained when Bob's choices of observables are  $B_{4,1} = (A_{4,1} +A_{4,2})/\sqrt{2}$ , $B_{4,2} = A^{\prime}_{4}$, $B_{4,3} = A''_{4}$ and $B_{4,4} = ({\tilde A}_{4,1} -{\tilde A}_{4,2})/\sqrt{2}$. Consequently, the optimal quantum success probability from Eq. (\ref{sp4-4}) is given by
			\begin{eqnarray}
   \label{bb}
		\left(\mathcal{P}_{4}^{g_4=4}\right)_{Q}^{opt} =\dfrac{1}{2}+\dfrac{2+\sqrt{2}}{8}\approx0.926.
			\end{eqnarray}
which is greater than the preparation noncontextual bound $0.875$. One of the possible choices of Alice's observables providing the optimal quantum value is the following. 
		\begin{eqnarray}
		\nonumber
		    &&A_{4,1} = \sigma_{x} \ \otimes \ \sigma_{x} \ \otimes \ \sigma_{x} ,\  \  {\tilde A}_{4,1} = \mathbb{I} \ \otimes \ \sigma_{x} \ \otimes \ \mathbb{I} ;\\
		    &&A^{\prime}_{4} = \sigma_{x} \ \otimes \ \mathbb{I} \ \otimes \mathbb{I} ;\  \  A^{\prime\prime}_{4} =\mathbb{I} \ \otimes \mathbb{I} \ \otimes \ \sigma_{x} ,\\
		    \nonumber
		    &&A_{4,2}= \sigma_{x} \ \otimes \ \sigma_{z} \otimes \ \sigma_{x},\  \  {\tilde A}_{4,2} = \mathbb{I} \ \otimes \ \sigma_{z} \ \otimes \mathbb{I} ,
		\end{eqnarray}
  and Bob's observables can be fixed accordingly.

  \subsection{Optimal quantum advantage for $\mathbb{G}_{4,3}$ }
In the $4$-bit PORAC, the parity-set corresponding to  $\mathbb{G}_{4,3}$ contains five elements, $s=1110,1101,1011,0111$ and $1111$. The PO condition has to be satisfied for all of them. For such a parity-set, Alice is allowed to communicate two bits without revealing the parity information and consequently the preparation noncontextual bound is $(\mathcal{P}_{4}^{g_4=3})_{pnc}=3/4$. 

In quantum theory, Alice encodes her inputs $x\in \{0,1\}^4$  into sixteen quantum states  $\rho_x \in \mathbb{C}^{2} \otimes \mathbb{C}^{2}$ and sends to Bob. As explicitly shown in the Appendix \ref{optm43}, the quantum success probability can be cast as 
\begin{equation}\label{p43}
(\mathcal{P}_{4}^{g_4=3})_Q=\dfrac{1}{2}+\dfrac{\mathrm{tr}[\Delta_4^{g_4=3}]_Q}{128}
\end{equation}
where  $\Delta_4^{g_4=3}$ has the form
\begin{eqnarray}\label{g43}
\nonumber
	\Delta_4^{g_4=3}&=&\Big[(A_{4,1}+A_{4,2}+A_{4,3}+ A_{4,4})B_{4,1}\\ &+&(A_{4,1}-A_{4,2}+A_{4,3}- A_{4,4}))B_{4,2}\\ \nonumber
	&+& (A^{\prime}_{4,1}+A^{\prime}_{4,2}+ A^{\prime}_{4,3}+A^{\prime}_{4,4})B_{4,3}\\ \nonumber
 &+&(A^{\prime}_{4,1}-A^{\prime}_{4,2}- A^{\prime}_{4,3}+A^{\prime}_{4,4})B_{4,4}
 \Big]
\end{eqnarray}

For the element $s=1111$, parity conditions $\sum_{x|s.x=0}\rho_{x}=\sum_{x|s.x=1}\rho_{x}=\mathrm{I}\otimes\mathrm{I}$ is trivially satisfied. But for the elements, $s=1110,1101,1011$ and $0111$, four different nontrivial relations between Alice’s observables are derived as follows.
\begin{eqnarray}
\nonumber
	&&A_{4,1}^{\prime}+A_{4,2}^{\prime}+A_{4,3}-A_{4,4}^{\prime}=0; \ \
	A_{4,1}-A_{4,2}-A_{4,3}^{\prime}-A_{4,4}=0;\\
	\nonumber
	&&	A_{4,1}^{\prime}-A_{4,2}^{\prime}-A_{4,3}-A_{4,4}^{\prime}=0; \ \
	A_{4,1}+A_{4,2}-A_{4,3}^{\prime}+A_{4,4}=0\\
\end{eqnarray}

The above functional relations between Alice's observables impose the constraint relations 
\begin{equation}\label{nontrivial}
    A_{4,1}=A_{4,3}^{\prime};\  \ \ \ \ \ A_{4,1}^{\prime}=A_{4,4}^{\prime}.
\end{equation}

 Using Eq. (\ref{nontrivial}), the correlation function $\Delta_4^{g_4=3}$ in Eq. (\ref{g43}) can be re-written as 
\begin{eqnarray}	
&&\Delta_{4}^{g_4=3} =(A_{4,1}+A_{4,2}+A_{4,3}+A_{4,4}) B_{4,1}\\
\nonumber
&+& (A_{4,1}-A_{4,2}+A_{4,3}-A_{4,4}) B_{4,2}\\
\nonumber
&+& (2A'_{4,1}+A'_{4,2}+A_{4,1}) B_{4,3}+ (2A'_{4,1}-A'_{4,2}-A_{4,1}) B_{4,4}.
\end{eqnarray} 

It is then straightforward to check that $\mathrm{tr}[\Delta_{4}^{g_4=3}]_{pnc}=32$, leading to $\left(\mathcal{P}_4^{g_4=3}\right)_{pnc}=3/4$. The optimal quantum success probability is derived  as $(\mathcal{P}_{4}^{g_4=3})_Q\approx 0.819 > 3/4$, thereby revealing quantum advantage over preparation noncontextual bound. The details of the derivation is placed in Appendix \ref{optm43}.
					
\section{$n$-bit generalized quantum PORAC}	
In an $n$-bit PORAC the generalized parity-set is $\mathbb{G}_{n,g_n}=\{x|x \in \{0,1\}^{n} , \sum_{r}x_{r} \geq g_n\}$. We demonstrate the quantum advantage for the case $g_n=n$. For such a case only one parity element $s=111....11$, is allowed such that $(n-1)$ bits can be communicated to Bob without revealing the parity information. 

Alice encodes her inputs $\{0,1\}^{n}$ into $2^{n}$ quantum states $\rho_{x}$ following the same strategy as showcased for $3$ and $4$-bit PORAC. Note here that, for such encoding in quantum theory, the PO condition $\sum_{x|s.x=0}\rho_{x}=\sum_{x|s.x=1}\rho_{x}=\mathrm{I}^{\otimes (n-1)}$ is satisfied. Following the same argument as for $3$ and $4$-bit case, Alice's generalised observables can be constructed.

Bob's decoding strategy remains same as earlier who performs projective measurements $ \{\Pi_{n,y}^0,\Pi_{n,y}^1\}$ with $y=1,2,...,n$. Using Eq. (\ref{QSP}), the success probability in quantum theory for $n$-bit PORAC with $g_n=n$ can be written as 
			\begin{equation}
				(\mathcal{P}_{n}^{g_n=n})_Q=\dfrac{1}{2}+\dfrac{\mathrm{tr}[\Delta_{n}^{g_n=n}]_Q}{2^{n+1}n}.
			\end{equation}
			where the correlation function $\Delta_{n}^{g_n=n}$ is derived as
			\begin{eqnarray}
			\nonumber
				\Delta_{n}^{g_n=n}&=& (A_{n,1}+A_{n,2}) B_{n,1}+(\tilde{A}_{n,1}-\tilde{A}_{n,2}) B_{n,n}\\
				&+& 2\sum_{k=1}^{n-2} A_{n}^k  B_{n,k+1}.
			\end{eqnarray}
			The preparation noncontextual bound $\mathrm{tr}[\Delta_{n}^{g_n=n}]_{pnc}\leq 2^n(n-1)$ and consequently, $(\mathcal{P}_{n}^{g_n=n})_{pnc}\leq (2n-1)/2n$. 
   
   Given Bob's choices of observables as $B_{n,1} = (A_{n,1} +A_{n,2})/\sqrt{2}$ and $B_{n,n} = (\tilde{A}_{n,1} -\tilde{A}_{n,2})/\sqrt{2}$ ,but $B_{n,k+1} = A_{n}^{k}$  with $k\in\{1, ...,n-2 \}$, the optimal quantum success probability is derived as
			\begin{eqnarray}
				(\mathcal{P}_{n}^{g_n=n})_{Q}\leq\dfrac{1}{2}+\dfrac{(n-2)+\sqrt{2}}{2n}>(\mathcal{P}_{n}^{g_n=n})_{pnc}
			\end{eqnarray}
			thereby providing the quantum advantage for the generalized PORAC. 

   By using the observables required for $n = 3,4$ cases, we can recursively fix  Alice's
observables  for $n\geq5$ as follows.
\begin{eqnarray}
\nonumber
    A_{n,1}=A_{n-1,1} \otimes \sigma_{x},\  \  \ A_{n,2}=A_{n-1,2} \otimes \sigma_{x},\\
     A_{n}^k = \left \{
    \begin{aligned}
    &\mathbb{I}\otimes A_{n-1}^{k-1}, &&\  \  \text{for}\  \  k=n-2,\\
    &A_{n-1}^{k} \otimes \mathbb{I}, &&\  \  \text{for}\  \ \ k=1\  \  \text{to}\  \  n-3,
  \end{aligned} \right.\\
    \nonumber
    \tilde{A}_{n,1}=A_{n,1}\prod\limits_{k=1}^{n-2}{A}_{n}^{k}\  \   \text{and}\  \  \tilde{A}_{n,2}=A_{n,2}\prod\limits_{k=1}^{n-2}{A}_{n}^{k}.
\end{eqnarray}
Similarly, Bob's observables can also be fixed.

\section{Summary and Discussion} In summary, we proposed a nontrivial generalization of standard $n$-bit two-level PORAC by introducing a generalized parity-set $\mathbb{G}_{n, g_{n}}$ where $g_{n}\in \{2,3.. n\}$. Thus, for a given $n$ there exists $n-1$ different PORAC games. For $g_{n}=2$, the standard PORAC in \cite{spek09} can be recovered. As argued in \cite{spek09}, the PO condition in an operational theory is equivalently represented in an ontological model of the said operational theory satisfying preparation non-contextuality.  We explicitly derived the classical preparation noncontextual bound on the success probability in our generalized PORAC for any arbitrary $n$ and $g_n$. Further, we demonstrated the quantum supremacy over the preparation noncontextual model. 

By considering the prepare-and-measure version of PORAC, we explicitly derived the optimal quantum advantage for $3$-bit PORAC with $g_3=3$, and for $4$-bit cases with $g_4=3$ and $g_4=4$. For arbitrary $n\geq 5$ we provided the explicit derivation of optimal quantum success probability for the class of PORAC when $g_n=n$. However, following the procedure developed for $n=3$ and $4$,  quantum supremacy can be demonstrated for any arbitrary $g_n>2$ for any given $n$. This requires a lengthy analytical calculation or considerably high computational resources. This calls for further study. 

We conclude by stating a couple of open questions in connection to our work. Our generalized PORAC $\mathbb{G}_{n,g_n}$ with $n\geq 3$ and $g_n\geq 3$ can serve as device-independent dimension witness of Hilbert space. For example, in the  PORAC $\mathbb{G}_{3,3}$, the optimal quantum value of the success probability is obtained when Alice and Bob share a pair of two-qubit entangled states. If they use a two-qubit entangled state, then there will be an upper bound on the quantum success probability, and any value that exceeds that upper bound certifies the dimension of the Hilbert space.  This could be an interesting line of study. Our work can be further generalized for $d$-level PORAC \cite{hameedi17b}. Such a generalization could be an exciting avenue for future research.
\appendix

\begin{widetext}

\section{Proof of Theorem 1 for $n=3$} \label{app3}

Here we provide the detailed derivation of \emph{Theorem 1}. The proof is built upon the approach developed in \cite{spek09}. Let us first consider the $3$-bit PORAC with $g_3=3$ and provide a detailed derivation of the average success probability in a preparation noncontextual model  corresponding to the parity-set $\mathbb{G}_{3, 3}$. 

 By using Bayes' theorem, and by noting that the distribution over inputs $x$ is uniform, Eq. (\ref{eqq1}) in the main text can be expressed as a constraint on $p(m|P_{x})$ so that 
\begin{equation}
\label{eq2}\forall s \in \mathbb{G}_{3,3}\ : \sum_{x|x.s=0}p(m|P_{x})=\sum_{x|x.s=1}p(m|P_{x}).\end{equation}
Our aim is to show that $p(m|P_{x})$ can be expanded in the following form
\begin{equation}
\label{eq3}
p(m|P_{x})=p(0)\alpha^0(m)+\sum_{\substack{q_i,q_j=1 \\ q_i\neq q_j}}^{3}p(q_{i}q_{j})\sum_{l,l^{\prime}=0}^{1}\Big[\alpha^{1}_{q_i,l,q_j,l^{\prime}} \ \delta_{x_{q_{i}},l} \ \delta_{x_{q_{j}},l^{\prime}}\Big]
\end{equation}
where we lebel the Fourier coefficient $\alpha^{0}(m)$ for the all-zero string, and for the string with entry $1$ at position $q_{i}$ and/or $q_{j}$, we label the Fourier coefficient as $\alpha^{1}_{q_{i},q_{j}}(m)$. The quantity $p(q_{i}q_{j})$ signifies the normalised probability distribution over $q_{i},q_{j}\in \{0,...,3\}$ with $q_i\neq q_j$. If $q_i=q_j=0$, Alice sends a message from the distribution function $\alpha^{0}(m)$. On the other hand, upon obtaining $q_i(q_j) \in_{\substack{q_i\neq q_j}} \{1,2,3\}$, she sends the message from the distributions $\alpha^{1}_{q_i,l,q_j,l^{\prime}}$ with $l,l^{\prime}=\{0, 1\}$  depending on	the values of the $q_{i}^{th}$ and $q_{j}^{th}$ bits of $x$.
						
In optimal decoding, Bob gets no information about $x$ if $q_{i},q_{j} = 0$, hence $p(0) = 0$. In order to optimize the amount of information, Bob should be able to distinguish the distributions $\alpha^{1}_{q_{i},l,q_{j},l^{\prime}}$, i.e., their mutual supports need to be disjoint. Given the message $m$, Bob must determine from which	distribution it was sampled. Therefore, when $q_{i} = y$ and $q_{j}=y^{\prime}$ where $y\neq y^{\prime}$ with $y,y^{\prime}\in \{1,2,3\}$, Bob always finds winning condition $b = x_{q_{i}}$ and $b=x_{q_{j}}$ respectively. However, if either $q_{i}\neq y$ or $q_{j}\neq y^{\prime}$, or both, Bob knows nothing about $x_{q_{i}}$ or $x_{q_{j}}$ and thus he only randomly guesses the bit with probability $1/2$. 

For the parity-set $\mathbb{G}_{3, 3}$, let Alice and Bob agree beforehand that Alice will communicate the first $2$-bits. Then, for $y,y^{\prime}=\{1,2\}$, each occurring with probability $1/3$, Bob can predict the outcome with certainty. For the third bit, Bob can only randomly guess it with the probability of $1/2$. The average success probability in a preparation noncontextual model can then be derived as $(\mathcal{P}_{3}^{3})_{pnc}=\frac{1}{3}\times 2 +\frac{1}{3}\times \frac{1}{2}=\frac{5}{6}$. Instead of the first $2$-bits, Alice can send any two random bits, but the success probability remains unchanged. 

To prove Alice can at most send two bits of information, we now provide the detailed derivation of Eq. (\ref{eq3}). Let us consider $\delta \in \{0,1\}^3$ and define a function $\eta_\delta(x)=(-1)^{x.\delta}$ which maps $\{0,1\}^3$ to $\{1,-1\}$. The function obeys $\sum_{x}\bar{\eta}_{\delta^{\prime}}(x)\eta_\delta(x)=2^3\delta_{\delta^{\prime},\delta}$, where $\{\eta_\delta(x)\}$ forms a complete set of orthogonal vectors. Using Fourier expansion of $p(m|P_{x})$, we can write
\begin{equation}
	\label{fo}
p(m|P_{x})=\sum_{\delta\in \{0,1\}^3}\alpha_{\delta}(m)\eta_\delta(x)
\end{equation}
where $\alpha_{\delta}(m)$ are the Fourier coefficients can be written as \begin{equation}
\alpha_{\delta}(m)=\frac{1}{2^3}\sum_{x \in \{0,1\}^3}\left[\sum_{x|x.s=0}p(m|P_x)-\sum_{x|x.s=1}p(m|P_x)\right].
\end{equation}
Combining the parity-oblivous condition in Eq. (\ref{eq2}), we find 
\begin{equation}
\forall s \ \in \ \mathbb{G}_{3,3}:\alpha_{s}(m)=0.
\end{equation}
The parity-set $\mathbb{G}_{3,3}$ contain only one element $s=111$. Now, the strings with hamming weight $0$ to $2$ in turn leads to strings $r$ for which $\alpha_{\delta}(m)\neq0$. Among them, one is the all-zero string. For the rest of the  strings, $1$ appears at position $q_{i},q_{j} \in \{1, . . . , 3\}$ with $q_{i}\neq q_{j}$ and all other entries of $\delta$ are zero. For the all-zero string we lebel the Fourier coefficient $\alpha^{0}(m)$, and for the string with entry $1$ at position $q_{i}$ and/or $q_{j}$, we label the Fourier coefficient as $\alpha^{1}_{q_{i},q_{j}}(m)$. Hence, under the parity-oblivious constraint, Eq. (\ref{fo}) reduces to
\begin{eqnarray}
\label{eq4}
	p(m|P_{x})=\alpha^{0}(m)+\sum_{\substack{q_i,q_j=1 \\ {q_i}\neq {q_j}}}^{3}\alpha^{1}_{q_i,q_j}(m)(-1)^{x_{q_i}+x_{q_j}}
\end{eqnarray}
	where $\alpha^{1}_{{q_i},{q_j}}=\alpha^{1}_{{q_i}} \alpha^{1}_{{q_j}}$ as  $q_i^{\text{th}}$ bit and $q_j^{\text{th}}$ bit are independent in the bit string $x$. 
	
	Next, inserting  $(-1)^{x_{q_i}+x_{q_j}}=\sum_{l^{\prime},l=0}^{1}(-1)^{l+ l^{\prime}}\delta_{x_{q_i},l}\delta_{x_{q_j},l^{\prime}}$ and $1=\sum_{l^{\prime},l=0}^{1}\delta_{x_{q_i},l}\delta_{x_{q_j},l^{\prime}}$ into Eq. (\ref{eq4}), we can write
	\begin{eqnarray}
	\nonumber
	\label{eqq5}
	p(m|P_{x})=a_0(m)&+&
	\sum_{\substack{q_i,q_j=1 \\ q_i\neq q_j}}^{3}\Big[a_{q_i,0,q_j,0}\delta_{x_{q_i},0}\delta_{x_{q_j},0}+a_{q_i,0,q_j,1}\delta_{x_{q_i},0}\delta_{x_{{q_j}},1}\\
	&+&a_{{q_i},1,{q_j},0}\delta_{x_{q_i},1}\delta_{x_{{q_j}},0}+a_{{q_i},1,{q_j},1}\delta_{x_{q_i},1}\delta_{x_{q_j},1}\Big].
	\end{eqnarray}
	where we have defined nonnegative coefficients as
	\begin{eqnarray}
	\nonumber
	a_{q_{i},0,{q_{j}},0}=2\alpha^{1}_{q_{i},0,q_{j},0}(m),\ \ &&\text{if sgn}(\alpha^{1}_{q_{i},0}(m))\geq 0\  , \text{ sgn}(\alpha^{1}_{q_{j},0}(m))\geq 0\\ 
	\nonumber
	a_{q_{i},1,{q_{j},1}}=2\alpha^{1}_{q_{i},1,q_{j},1}(m),\ \ &&\text{if sgn}(\alpha^{1}_{q_{i},1}(m))< 0\  , \text{ sgn}(\alpha^{1}_{q_{j},1}(m))< 0\\
	\nonumber
	a_{q_{i},0,{q_{j}},1}=-2\alpha^{1}_{q_{i},0,q_{j},1}(m),\ \ &&\text{if sgn}(\alpha^{1}_{q_{i},0}(m))\geq 0\  , \text{ sgn}(\alpha^{1}_{q_{j},1}(m))< 0\\
	\nonumber
	a_{q_{i},1,{q_{j}},0}=-2\alpha^{1}_{q_{i},1,q_{j},0}(m),\ \ &&\text{if sgn}(\alpha^{1}_{q_{i},1}(m))< 0\  , \text{ sgn}(\alpha^{1}_{q_{j},0}(m))\geq 0.\\
	\end{eqnarray}

	It remains to prove that the coefficient $a_0(m)$ is also positive semi-definite. For this, we
	define a $3$-bit string $z(m)$ as 
\begin{equation}
z_{q_{i}}(m)z_{q_{j}}(m)\equiv
\begin{cases}
00\  \  \  \  \text{if sgn}(\alpha^{1}_{q_{i},0}(m))\geq 0\  , \text{ sgn}(\alpha^{1}_{q_{j},0}(m))\geq 0;\\ 
11\  \  \  \  \text{if sgn}(\alpha^{1}_{q_{i},1}(m))< 0\  , \text{ sgn}(\alpha^{1}_{q_{j},1}(m))< 0; \ \  \      \    \   \\ 
01\  \  \  \  \text{if sgn}(\alpha^{1}_{q_{i},0}(m))\geq 0\  , \text{ sgn}(\alpha^{1}_{q_{j},1}(m))< 0;\\
10\  \  \  \  \text{if sgn}(\alpha^{1}_{q_{i},1}(m))< 0\  , \text{ sgn}(\alpha^{1}_{q_{j},0}(m))\geq 0.
	\end{cases}
	\end{equation}
	where $z_{q_{i}}(m)z_{q_{j}}(m)$ encodes the signs of the Fourier coefficients.	It follows from this definition that
	\begin{eqnarray}
	\nonumber
	a_{q_{i},0,q_{j},0}\delta_{z_{q_{i}}(m),0}\delta_{z_{q_{j}}(m),0}+a_{q_{i},0,q_{j},1}\delta_{z_{q_{i}}(m),0}\delta_{z_{q_{j}}(m),1}+a_{q_{i},1,q_{j},0}\delta_{z_{q_{i}}(m),1}\delta_{z_{q_{j}}(m),0}+a_{q_{i},1,q_{j},1}\delta_{z_{q_{i}}(m),1}\delta_{z_{q_{j}}(m),1}=0\\
	\end{eqnarray}
	and consequently 
	\begin{equation}
	\forall\ q_{i},\ q_{j},\ \  \ p(m|P_{z(m)})=a_{0}(m),
	\end{equation}
	which establishes that $a_{0}(m)\geq 0$. Using normalization of $p(m|P_{x})$ together with Eq. (\ref{eqq5}) we obtain
	\begin{equation}
	\forall\ x, \  \sum_{m}p(m|P_{x})=\sum_{m}a_{0}(m)+\sum_{\substack{q_{i},q_{j}=1 \\ q_{i}\neq q_{j}}}^{3}\sum_{m}a_{q_{i},x_{q_{i}},q_{j},x_{q_{j}}}(m)=1,
	\end{equation}
	
	Noting the fact that  $\sum_{m}a_{{q_{i}},x_{q_{i}}q_{j},x_{q_{j}}}$ is independent of $x_{q_{i}}$ and $x_{q_{j}}$, i.e., $\sum_{m}a_{{q_{i}},x_{q_{i}}q_{j},x_{q_{j}}}=\sum_{m}a_{q_{j},{q_{j}}}(m)$, and   we introduce the following re-labeling.
	
	\begin{eqnarray}
	\nonumber
	&&p(0)=\sum_{m}a_0(m), \ \ \  \ p(q_{i},q_{j})=\sum_{m}a_{q_{j},{q_{j}}}(m), \ \  \ \ \\
 	&&	\alpha^0(m)=\frac{a_0(m)}{p(0)},\ \ \  \ \alpha^{1}_{q_{i},l,q_{j},l^{\prime}}(m)=\frac{a_{q_{i},l,q_{j},l^{\prime}}(m)}{p(q_{i},q_{j})}
\end{eqnarray}
	Here, $p(q_{i},q_{j})$ is the probability distribution over $q_{i},q_{j} \in \{0, . . . , 3\}$. $\alpha^0(m)$ is the probability distribution over $m$ and for each $q_{i},q_{j} \in \{0, . . . , 3\}$ and $l,l^{\prime} \in \{0,1\}$, $\alpha^{1}_{q_{i},l,q_{j},l^{\prime}}(m)$ is a probability distribution over $m$. Hence, we can
	write Eq. (\ref{eqq5}) as
	\begin{equation}
	p(m|P_{x})	=p(0)\alpha^{0}(m)+	\sum_{\substack{q_{i},q_{j}=1 \\ q_{i}\neq q_{j}}}^{3}p(q_{i}q_{j})\sum_{l,l^{\prime}=0}^{1}\alpha^{1}_{q_{i},l,q_{j},l^{\prime}}(m)\delta_{x_{q_{i}},l}\delta_{x_{q_{j}},l^{\prime}}
\end{equation}
	which is the Eq. (\ref{eq3}) that we promised to prove.
\section{Proof of Theorem 1 for arbitrary $n$}
\label{proofmn}
For $n$-bit PORAC, Alice holds $x= x_{1}x_{2}... x_{n}\in \{0,1\}^{n}$ with $p_{A}(x) = 1/2^{n}$, and Bob holds $y \in \{1,2,...,n\} $ with
$p_{B}(y) = 1/n$. Parity-oblivious condition demands that the classical message $m$ sent from Alice to Bob must satisfy
\begin{equation}
	\label{eq11}
	\forall s \in \mathbb{G}_{n,g_n}\ : \sum_{x|x.s=0}p(P_{x}|m)=\sum_{x|x.s=1}p(P_{x}|m)
\end{equation}
where $g_n=\{2,3,...,n\}$ for any arbitrary $n$.

By employing Bayes' theorem, and considering that the distribution over inputs $x$ is uniform, Eq. (\ref{eq11}) can be expressed as a constraint on $p(m|P_{x})$ as
\begin{equation}
	\label{eq12}
	\forall s  \in \mathbb{G}_{n,g_n}\ : \sum_{x|x.s=0}p(m|P_{x})=\sum_{x|x.s=1}p(m|P_{x}).
\end{equation}
The expression  $p(m|P_{x})$ implies 
\begin{equation}
	\label{eq13}
	p(m|P_{x})=p(0)\alpha^0(m)+
	\sum_{q=1}^{n}p(q)\prod_{q=1}^{n}\sum_{l_{q}=0}^{1}\Big[\alpha^{1}_{q,l_{q}}\delta_{x_{q},l_{q}}\Big]
\end{equation}
where $p(q)$ signifies the normalised probability distribution over $\{0,...,n\}$ with $q=(q_1,...,q_{g_n-1})$. If $\forall n, \ q(q_1,...,q_{g_n-1})=0$, Alice sends a message from the distribution function $\alpha^0(m)$, otherwise she sends from the distributions $\alpha^{1}_{q,l_{q}}$ upon obtaining $q \in \{1,...,n\}$ where $l_{q}=0 \ \text{or}\ 1$.

Following the similar steps adopted for $g_3=3$ case, Fourier coefficients $\alpha_{\delta}(m)$ can be written as
\begin{equation}
	\alpha_{\delta}(m)=\frac{1}{2^n}\sum_{x \in \{0,1\}^n}\Big[\sum_{x|x.s=0}p(m|P_x)-\sum_{x|x.s=1}p(m|P_x)\Big].
\end{equation}
also we have 
\begin{equation}
	\forall s \ \in \ \mathbb{G}_{n,g_n}:\alpha_{s}(m)=0.
\end{equation}
The strings with hamming weight $0$ to $g_n-1$ in turn leads to strings $\delta$ for which $\alpha_{\delta}(m)\neq0$. Among them, one is the all-zero string for which we denote the Fourier coefficient as $\alpha^{0}(m)$. For the rest of the strings, $1$ appears at position $q \in \{1, . . . , n\}$ with all other entries of $\delta$ are zero. In such cases, we label the Fourier coeffient as $\alpha^{1}_{q}(m)$. Hence, under the parity-oblivious constraint, $p(m|P_{x})$ reduces to
\begin{equation}
	\label{eq14}
	p(m|P_{x})=\alpha^{0}(m)+\sum_{q=1}^{n}\alpha^{1}_{q}(m)(-1)^{\oplus_{2}x_{q}}
\end{equation}
where in $\alpha^{1}_{q}=\prod_{q=1}^{n}\alpha^{1}_{q}$, as all bits are independent of the bit string $x$.

Now, insurting  $(-1)^{\oplus_{2}x_{q}}=\sum_{l_{q}=0}^{1}(-1)^{\oplus_{2}l_{q}}\delta_{x_{q},l_{q}}$ and $1=\sum_{l_{q}=0}^{1}\delta_{x_{q},l_{q}}$ into Eq. (\ref{eq14}), we can write the expression as
\begin{equation}
	\label{eq15}
	p(m|P_{x})=a_0(m)+
	\prod_{q=1}^{n}\sum_{l_{q}=0}^{1}\Big[a_{q,l_{q}}\delta_{x_{q},l_{q}}\Big].
\end{equation}
where we have defined nonnegative coefficients as
\begin{eqnarray}
	a_{q,l_{q}}=(-1)^{l_{q}}2\alpha^{1}_{q,l_{q}}(m)
\end{eqnarray}
depending on the values of $\text{sgn}(\alpha^{1}_{q,l_{q}}(m))\geq 0\ $, or $\text{sgn}(\alpha^{1}_{q,l_{q}}(m))< 0$, and $l_{q}\in \{0,1\}$.\\

We have implicitly defined a constant $a_0(m)$, where it remains to prove that $a_0(m)$ is positive semidefinite. To prove $a_0(m)$ as nonnegative, let a $n$-bit string $z(m)$ defined as
\begin{eqnarray}
	 z_{q_1}(m)z_{q_2}(m)...z_{q_{g_n-1}}(m)\in  \{0,1\}^{g_n-1}\ \ \ \text{if sgn}(\alpha^{1}_{q,l_{q}=0}(m))\geq 0, \  \   \text{ sgn}(\alpha^{1}_{q,l_{q}=1}(m))< 0,	
\end{eqnarray}
where $z_{q}(m)$ encodes the signs of the Fourier coefficients. Then it follows from this definition that
\begin{equation}
	\prod_{q=1}^{n}\sum_{l_{q}=0}^{1}\Big[a_{q,l_{q}}\delta_{x_{q},l_{q}}\Big]=0
\end{equation}
and consequently that
\begin{equation}
	\forall\ q,\ \  \ p(m|P_{z(m)})=a_{0}(m),
\end{equation}
which establishes that $a_{0}(m)\geq 0$.

Using normalization together with Eq. (\ref{eq15}) to obtain:
\begin{equation}
	\forall\ x, \  \sum_{m}p(m|P_{x})=\sum_{m}a_{0}(m)+\prod_{q=1}^{n}\sum_{l_{q}=0}^{1}\sum_{m}a_{q,l_{q}}\delta_{x_{q},l_{q}}(m)=1,
\end{equation}

Introducing the re-labeling as 

\begin{eqnarray}
\nonumber
	\label{re-l}
	p(0)=\sum_{m}a_0(m) ,\   \    \   \   \   \    p(q)=\sum_{m}a_{q}(m); \   \    \   \   \   \ \alpha^0(m)=\frac{a_0(m)}{p(0)}, \ \ \alpha^{1}_{q,l_{q}}(m)=\frac{a_{q,l_{q}}(m)}{p(q)}.
\end{eqnarray}
here, $p(q)$ can be interpreted as a probability distribution over $q \in \{0, . . . , n\}$, $\alpha^0(m)$ is a probability distribution over $m$ when $q \equiv (q_1, . . . , q_{g_n-1})=0$. Also for each $q \equiv (q_1, . . . , q_{g_n-1}) \in \{1, . . . , n\}$ (with $q_{k}\neq q_{k^{\prime}}$, and $k,k^{\prime}=\{1,2,...,n\}$) and $l_{q} \in \{0,1\}$, $\alpha^{1}_{q,l_{q}}(m)$ is a probability distribution over $m$. Hence, using Eq. (\ref{re-l}) we can
write Eq. (\ref{eq15}) as
\begin{equation}
	p(m|P_{x})=p(0)\alpha^0(m)+
	\sum_{q=1}^{n}p(q)\prod_{q=1}^{n}\sum_{l_{q}=0}^{1}\alpha^{1}_{q,l_{q}}\delta_{x_{q},l_{q}}.	
\end{equation}
which is the Eq. (\ref{eq13}).

Thus, the parity-oblivious task can be interpreted as follows: Alice samples $q$ from $p(q)$, if she obtains $q = 0$ she sends a message sampled from $\alpha^{0}(m)$ whereas if she obtains $q \in \{1, . . . , n\},$ she sends a message distributions from $\alpha^{1}_{q,l_{q}}$ depending on
the value of the $q^{\text{th}}$ bit of $x$. To achieve the optimal value of $\mathcal{P}_{n}$ in Eq. (1), Bob gets no information about $x$
if $q = 0$, Hence $p(0) = 0$. In order to optimize the amount of information the distributions $\alpha^{1}_{q,l_{q}}$ have to be distinguishable for Bob. Given any value of $m$, Bob can certainly determine from which
distribution it was sampled, i.e., their mutual supports need to be disjoint. Therefore, when $q = y$, Bob always finds $b = x_q$. However, if $q\neq y$ Bob knows nothing
about $x_q$ and thus only randomly guesses the bit with probability $1/2$. 

For the parity-set $\mathbb{G}_{n,g_n}$, let Alice and Bob agreed beforehand that Alice will reveal the first $(g_n-1)$-bits. For $y=\{1,...,g_n-1\}$, each occurring with probability $1/n$, Bob can predict the outcome with certainty. And, for rest of $(n-g_n+1)$ bits, Bob can only randomly guess the bit with probability $1/2$. The average value of $(\mathcal{P}_{n}^{g_n})_{pnc}$ is then
\begin{align}
	(\mathcal{P}_{n}^{g_n})_{pnc}=(g_n-1)\times\frac{1}{n}+(n-g_n+1)\times\frac{1}{2}=\frac{(g_n+n-1)}{2n}.
\end{align}
which is claimed in Eq. (\ref{gpnc}). 

\section{Detailed derivation of optimal quantum success probability for $g_4=4$ }
\label{optm44}

In the $4$-bit quantum PORAC game for the parity-set $\mathbb{G}_{4,4}$, Alice encodes her inputs into sixteen quantum states $\rho_{x}$ and sends it to Bob.
The decoding measurement for Bob is $ \{\Pi_{4,y}^{0},\Pi_{4,y}^{1}\}$ with $y=1,2,3,4$. The quantum success probability using Eq. (\ref{QSP}) from the main text can be explicitly written as
\begin{eqnarray}
	\label{eq7}
	\nonumber
	(\mathcal{P}_4^{g_4=4})_{Q}=\dfrac{1}{2}+\dfrac{1}{64}&\mathrm{tr}&[\rho_{0000}(\Pi_{4,1}^{0}+\Pi_{4,2}^{0}+\Pi_{4,3}^{0}+\Pi_{4,4}^{0})-\rho_{1111}(\Pi_{4,1}^{0}+\Pi_{4,2}^{0}+\Pi_{4,3}^{0}+\Pi_{4,4}^{0})+\rho_{0011}(\Pi_{4,1}^{0}+\Pi_{4,2}^{0}-\Pi_{4,3}^{0}-\Pi_{4,4}^{0})\\
	\nonumber
	&-&\rho_{1100}(\Pi_{4,1}^{0}+\Pi_{4,2}^{0}-\Pi_{4,3}^{0}-\Pi_{4,4}^{0})+\rho_{0101}(\Pi_{4,1}^{0}-\Pi_{4,2}^{0}+\Pi_{4,3}^{0}-\Pi_{4,4}^{0})-\rho_{1010}(\Pi_{4,1}^{0}-\Pi_{4,2}^{0}+\Pi_{4,3}^{0}-\Pi_{4,4}^{0})\\
	\nonumber
	&+&\rho_{0110}(\Pi_{4,1}^{0}-\Pi_{4,2}^{0}-\Pi_{4,3}^{0}+\Pi_{4,4}^{0})-\rho_{1001}(\Pi_{4,1}^{0}-\Pi_{4,2}^{0}-\Pi_{4,3}^{0}+\Pi_{4,4}^{0})+\rho_{0001}(\Pi_{4,1}^{0}+\Pi_{4,2}^{0}+\Pi_{4,3}^{0}-\Pi_{4,4}^{0})\\
	\nonumber
	&-&\rho_{1110}(\Pi_{4,1}^{0}+\Pi_{4,2}^{0}+\Pi_{4,3}^{0}-\Pi_{4,4}^{0})+\rho_{0010}(\Pi_{4,1}^{0}+\Pi_{4,2}^{0}-\Pi_{4,3}^{0}+\Pi_{4,4}^{0})-\rho_{1101}(\Pi_{4,1}^{0}+\Pi_{4,2}^{0}-\Pi_{4,3}^{0}+\Pi_{4,4}^{0})\\
	\nonumber
	&+&\rho_{0100}(\Pi_{4,1}^{0}-\Pi_{4,2}^{0}+\Pi_{4,3}^{0}+\Pi_{4,4}^{0})-\rho_{1011}(\Pi_{4,1}^{0}-\Pi_{4,2}^{0}+\Pi_{4,3}^{0}+\Pi_{4,4}^{0})+\rho_{0111}(\Pi_{4,1}^{0}-\Pi_{4,2}^{0}-\Pi_{4,3}^{0}-\Pi_{4,4}^{0})\\
	&-&\rho_{1000}(\Pi_{4,1}^{0}-\Pi_{4,2}^{0}-\Pi_{4,3}^{0}-\Pi_{4,4}^{0})] .
\end{eqnarray}
which can further be re-written as
\begin{eqnarray}
	\label{B5}
	\nonumber
	(\mathcal{P}_4^{g_4=4})_{Q}=\dfrac{1}{2}+\dfrac{1}{128}&\mathrm{tr}&[(\rho_{0000}-\rho_{1111}+\rho_{0011}-\rho_{1100}+\rho_{0101}-\rho_{1010}+\rho_{0110}-\rho_{1001})B_{4,1}\\
	\nonumber
	&+&(\rho_{0000}-\rho_{1111}+\rho_{0011}-\rho_{1100}-\rho_{0101}+\rho_{1010}-\rho_{0110}+\rho_{1001})B_{4,2}\\
	\nonumber
	&+&
	(\rho_{0000}-\rho_{1111}-\rho_{0011}+\rho_{1100}+\rho_{0101}-\rho_{1010}-\rho_{0110}+\rho_{1001})B_{4,3}\\
	\nonumber
	&+&
	(\rho_{0000}-\rho_{1111}-\rho_{0011}+\rho_{1100}-\rho_{0101}+\rho_{1010}+\rho_{0110}-\rho_{1001})B_{4,4}\\
	&+&
	(\rho_{0001}-\rho_{1110}+\rho_{0010}-\rho_{1101}+\rho_{0100}-\rho_{1011}+\rho_{0111}-\rho_{1000})B_{4,1}\\
	\nonumber
	&+&(\rho_{0001}-\rho_{1110}+\rho_{0010}-\rho_{1101}-\rho_{0100}+\rho_{1011}-\rho_{0111}+\rho_{1000})B_{4,2}\\
	\nonumber
	&+&
	(\rho_{0001}-\rho_{1110}-\rho_{0010}+\rho_{1101}+\rho_{0100}-\rho_{1011}-\rho_{0111}+\rho_{1000})B_{4,3}\\
	\nonumber
	&-&
	(\rho_{0001}-\rho_{1110}-\rho_{0010}+\rho_{1101}-\rho_{0100}+\rho_{1011}+\rho_{0111}-\rho_{1000})B_{4,4}]
\end{eqnarray}

Similar to the case of $\mathbb{G}_{3,3}$, to obtain optimal quantum success probability, we first analyze the term $Tr[(\rho_{0000}-\rho_{1111}+\rho_{0011}-\rho_{1100}+\rho_{0101}-\rho_{1010}+\rho_{0110}-\rho_{1001})B_{4,1}]$. Note that to get maximum value $\rho_{0000},\rho_{0011},\rho_{0101}$ and $\rho_{0110}$ has to be the eigenstates of $B_{4,1}$ with eigenvalue $+1$, and $\rho_{1100},\rho_{1010},\rho_{1001}$ and $\rho_{1111}$ has to be  eigenstates of $B_{4,1}$ with eigenvalue $-1$. Since we consider three-qubit system, $\rho_{x}$ with $s.x=0$ are orthogonal pure states implying  $\sum_{x|s.x=0}\rho_{x}=\mathrm{I}\otimes \mathrm{I}\otimes \mathrm{I}$. Similar arguments can be made for the second, third and forth term with $B_{4,2}$, $B_{4,3}$ and $B_{4,4}$. Consequently, the other four terms with $s.x=1$ will also follow same argument such that  $\rho_{0001},\rho_{1110},\rho_{0010},\rho_{1101},\rho_{0100},\rho_{1011},\rho_{0111}$ and $\rho_{1000}$ are also orthogonal pure states, i.e., $\sum_{x|s.x=1}\rho_{x}=\mathrm{I}\otimes \mathrm{I}\otimes \mathrm{I}$. This satisfies the PO condition. 

Hence, leading us to define three-qubit observables of the form 
\begin{eqnarray}
	\label{}
	\nonumber
	A_{4,1}&=&\rho_{0000}-\rho_{1111}+\rho_{0011}-\rho_{1100}+\rho_{0101}-\rho_{1010}+\rho_{0110}-\rho_{1001}\\
	\nonumber
	A^{\prime}_{4,1}&=&\rho_{0000}-\rho_{1111}+\rho_{0011}-\rho_{1100}-\rho_{0101}+\rho_{1010}-\rho_{0110}+\rho_{1001}\\
	\nonumber
	A^{\prime\prime}_{4,1}&=&
	\rho_{0000}-\rho_{1111}-\rho_{0011}+\rho_{1100}+\rho_{0101}-\rho_{1010}-\rho_{0110}+\rho_{1001}\\
	\nonumber
	\tilde{A}_{4,1}&=&
	\rho_{0000}-\rho_{1111}-\rho_{0011}+\rho_{1100}-\rho_{0101}+\rho_{1010}+\rho_{0110}-\rho_{1001}\\
	A_{4,2}&=&
	\rho_{0001}-\rho_{1110}+\rho_{0010}-\rho_{1101}+\rho_{0100}-\rho_{1011}+\rho_{0111}-\rho_{1000}\\
	\nonumber
	A^{\prime}_{4,2}&=&\rho_{0001}-\rho_{1110}+\rho_{0010}-\rho_{1101}-\rho_{0100}+\rho_{1011}-\rho_{0111}+\rho_{1000}\\
	\nonumber
	A^{\prime\prime}_{4,2}&=&
	\rho_{0001}-\rho_{1110}-\rho_{0010}+\rho_{1101}+\rho_{0100}-\rho_{1011}-\rho_{0111}+\rho_{1000}\\
	\nonumber
	\tilde{A}_{4,2}&=&\rho_{0001}-\rho_{1110}-\rho_{0010}+\rho_{1101}-\rho_{0100}+\rho_{1011}+\rho_{0111}-\rho_{1000}
\end{eqnarray}

Then the quantum success probability in Eq. (\ref{B5}) can be casted as
\begin{eqnarray}
	\label{B6}
	(\mathcal{P}_4^{g_4=4})_{Q}&=&\dfrac{1}{2}+\dfrac{\mathrm{tr}[\Delta_{4}^{g_4=4}]_Q}{128}.
\end{eqnarray}
where the function $\Delta_{4}^{g_4=4}$ is given by 
\begin{eqnarray}\label{B7}
				\Delta_4^{g_4=4}=(A_{4,1}+A_{4,2}) B_{4,1}+( \tilde{A}_{4,1}-\tilde{A}_{4,2}) B_{4,4}+ (A^{\prime}_{4,1}+A^{\prime}_{4,2}) B_{4,2}+ (A^{\prime\prime}_{4,1}+A^{\prime\prime}_{4,2}) B_{4,3}.
			\end{eqnarray}
Eq. (\ref{B6}) and Eq. (\ref{B7}) are the Eq. (\ref{sp4-4}) and Eq. (\ref{del44}) in the main text respectively .

\section{Detailed derivation of optimal quantum success probability for $4$-bit PORAC with $g_4=3$}
\label{optm43}

For $\mathbb{G}_{4,3}$, Alice prepares sixteen quantum states $\rho_{x}$ in quantum theory and communicates to Bob. Next, after receiving the states Bob performs projective measurements $ \{\Pi_{4,y}^{0},\Pi_{4,y}^{1}\}$ with $y=1,2,3,4$.
The quantum success probability using Eq. (\ref{QSP}) from the main text can be explicitly written as
\begin{eqnarray}
	\nonumber
	(\mathcal{P}_4^{g_4=3})_{Q}=\dfrac{1}{2}&+&\dfrac{1}{64}\mathrm{tr}[\rho_{0000}(\Pi_{4,1}^{0}+\Pi_{4,2}^{0}+\Pi_{4,3}^{0}+\Pi_{4,4}^{0})-\rho_{1111}(\Pi_{4,1}^{0}+\Pi_{4,2}^{0}+\Pi_{4,3}^{0}+\Pi_{4,4}^{0})\\
	\nonumber
	&+&\rho_{0011}(\Pi_{4,1}^{0}+\Pi_{4,2}^{0}-\Pi_{4,3}^{0}-\Pi_{4,4}^{0})-\rho_{1100}(\Pi_{4,1}^{0}+\Pi_{4,2}^{0}-\Pi_{4,3}^{0}-\Pi_{4,4}^{0})\\
	\nonumber
	&+&\rho_{0101}(\Pi_{4,1}^{0}-\Pi_{4,2}^{0}+\Pi_{4,3}^{0}-\Pi_{4,4}^{0})-\rho_{1010}(\Pi_{4,1}^{0}-\Pi_{4,2}^{0}+\Pi_{4,3}^{0}-\Pi_{4,4}^{0})\\
	\nonumber
	&+&\rho_{0110}(\Pi_{4,1}^{0}-\Pi_{4,2}^{0}-\Pi_{4,3}^{0}+\Pi_{4,4}^{0})-\rho_{1001}(\Pi_{4,1}^{0}-\Pi_{4,2}^{0}-\Pi_{4,3}^{0}+\Pi_{4,4}^{0})\\
	\nonumber
	&+&\rho_{0001}(\Pi_{4,1}^{0}+\Pi_{4,2}^{0}+\Pi_{4,3}^{0}-\Pi_{4,4}^{0})-\rho_{1110}(\Pi_{4,1}^{0}+\Pi_{4,2}^{0}+\Pi_{4,3}^{0}-\Pi_{4,4}^{0})\\
	\nonumber
	&+&\rho_{0010}(\Pi_{4,1}^{0}+\Pi_{4,2}^{0}-\Pi_{4,3}^{0}+\Pi_{4,4}^{0})-\rho_{1101}(\Pi_{4,1}^{0}+\Pi_{4,2}^{0}-\Pi_{4,3}^{0}+\Pi_{4,4}^{0})\\
	\nonumber
	&+&\rho_{0100}(\Pi_{4,1}^{0}-\Pi_{4,2}^{0}+\Pi_{4,3}^{0}+\Pi_{4,4}^{0})-\rho_{1011}(\Pi_{4,1}^{0}-\Pi_{4,2}^{0}+\Pi_{4,3}^{0}+\Pi_{4,4}^{0})\\
	&+&\rho_{0111}(\Pi_{4,1}^{0}-\Pi_{4,2}^{0}-\Pi_{4,3}^{0}-\Pi_{4,4}^{0})-\rho_{1000}(\Pi_{4,1}^{0}-\Pi_{4,2}^{0}-\Pi_{4,3}^{0}-\Pi_{4,4}^{0})] .
	\label{eq7}
\end{eqnarray}
Considering $\Pi_{4,y}^{0}=(\textbf{1}+B_{4,y})/2$, Eq. (\ref{eq7}) can be written as
\begin{eqnarray}
	\nonumber
	(\mathcal{P}_4^{g_4=3})_{Q}=\dfrac{1}{2}&+&\dfrac{1}{64\times 2}\mathrm{tr}[(\rho_{0000}-\rho_{1111}+\rho_{0011}-\rho_{1100})(B_{4,1}+B_{4,2})+(\rho_{0000}-\rho_{1111}-\rho_{0011}+\rho_{1100})(B_{4,3}+B_{4,4})\\
	\nonumber
	&+&(\rho_{0101}-\rho_{1010}+\rho_{0110}-\rho_{1001})(B_{4,1}-B_{4,2})+(\rho_{0101}-\rho_{1010}-\rho_{0110}+\rho_{1001})(B_{4,3}-B_{4,4})\\
	\nonumber
	&+&(\rho_{0001}-\rho_{1110}+\rho_{0010}-\rho_{1101})(B_{4,1}+B_{4,2})+(\rho_{0001}-\rho_{1110}-\rho_{0010}+\rho_{1101})(B_{4,3}-B_{4,4})\\
	&+&(\rho_{0100}-\rho_{1011}+\rho_{0111}-\rho_{1000})(B_{4,1}-B_{4,2})+(\rho_{0100}-\rho_{1011}-\rho_{0111}+\rho_{1000})(B_{4,3}
 +B_{4,4})] .
\end{eqnarray}

Similar to the $\mathbb{G}_{3,3}$ case, let us first analyze the term $\mathrm{tr}[(\rho_{0000}-\rho_{1111}+\rho_{0011}-\rho_{1100}) B_{4,1}]$. To get maximum value  $\rho_{0000}$ and $\rho_{0011}$ has to be the eigenstates of $B_{4,1}$ with eigenvalue $+1$, and $\rho_{1111}$ and $\rho_{1100}$ has to be  eigenstates of $B_{4,1}$ with eigenvalue $-1$. Since we consider two-qubit system, $\rho_{0000}, \rho_{0011}, \rho_{1110}$ and $\rho_{1111}$  are orthogonal pure states. Similar argument can be made for the second, third and forth terms. Now, if we consider $\mathrm{tr}[(\rho_{0001}-\rho_{1110}+\rho_{0010}-\rho_{1101})B_{4,1}]$, maximization criteria gives $\rho_{0001}, \rho_{0010},\rho_{1110}$ and $\rho_{1101}$ are also orthogonal pure states. The same argument holds for last three terms. This leads us to define two-qubit observables of the form
\begin{eqnarray}
	\label{obserg43}
	\nonumber
	A_{4,1}&=&\rho_{0000}-\rho_{1111}+\rho_{0011}-\rho_{1100}; \ \ \ 
	A_{4,3}=
	\rho_{0001}-\rho_{1110}+\rho_{0010}-\rho_{1101}\\
	\nonumber
	A^{\prime}_{4,1}&=&\rho_{0000}-\rho_{1111}-\rho_{0011}+\rho_{1100}; \ \ \ A^{\prime}_{4,3}=\rho_{0001}-\rho_{1110}-\rho_{0010}+\rho_{1101}\\
	A_{4,2}&=&
	\rho_{0101}-\rho_{1010}+\rho_{0110}-\rho_{1001};\ \ \ 
 A_{4,4}=
	\rho_{0100}-\rho_{1011}+\rho_{0111}-\rho_{1000}\\
	\nonumber
	{A}^{\prime}_{4,2}&=&
	\rho_{0101}-\rho_{1010}-\rho_{0110}+\rho_{1001};\ \ \ 
	{A}^{\prime}_{4,4}=\rho_{0100}-\rho_{1011}-\rho_{0111}+\rho_{1000}
\end{eqnarray}
satisfying $[A_{4,1},A^{\prime}_{4,1}]=[A_{4,2},A^{\prime}_{4,2}]=[A_{4,3},A^{\prime}_{4,3}]=[A_{4,4},A^{\prime}_{4,4}]=0$.

Thus, the quantum success probability can be written in terms of observables as
\begin{eqnarray}
	\nonumber
	(\mathcal{P}_4^{m_{4}=3})_{Q} =\dfrac{1}{2}+\dfrac{1}{128}&\Big[&(A_{4,1}+A_{4,2}+A_{4,3}+ A_{4,4})B_{4,1}+(A_{4,1}-A_{4,2}+A_{4,3}- A_{4,4})B_{4,2}\\ 
	&+& (A^{\prime}_{4,1}+A^{\prime}_{4,2}+ A^{\prime}_{4,3}+A^{\prime}_{4,4})B_{4,3}+(A^{\prime}_{4,1}-A^{\prime}_{4,2}- A^{\prime}_{4,3}+A^{\prime}_{4,4})B_{4,4}
 \Big]
\end{eqnarray} 

Since, Alice's observables can be represented in terms of quantum states, we can formulate an encoding scheme solely from the construction of the observables as follows. Let Alice, performs sequential non-selective degeneracy-breaking measurement on her set of two commuting observables $A_{4,1}$ and $A^{\prime}_{4,1}$ randomly. Then the quantum states $\rho_{0000},\rho_{1111},\rho_{0011},$ and $\rho_{1100}$ can be written in terms of Alice's observables in the following way,
\begin{eqnarray}\label{even431}
\nonumber
    \rho_{0000}&=&\frac{1}{4}(\mathrm{I}+A_{4,1}+A^{\prime}_{4,1}+\tilde{A}_{4,1}); \ \ \ 
    \rho_{0011}=\frac{1}{4}(\mathrm{I}+A_{4,1}-A^{\prime}_{4,1}-\tilde{A}_{4,1})\\
    \rho_{1100}&=&\frac{1}{4}(\mathrm{I}-A_{4,1}+A^{\prime}_{4,1}-\tilde{A}_{4,1}); \ \ \ 
    \rho_{1111}=\frac{1}{4}(\mathrm{I}-A_{4,1}-A^{\prime}_{4,1}+\tilde{A}_{4,1})
\end{eqnarray}
where $\tilde{A}_{4,1}$ occurs due to the existence of three mutually commuting observables two-qubit system.   

Similarly, using the commuting set of $A_{4,2}$ and $A^{\prime}_{4,2}$, other four quantum states $\rho_{1010},\rho_{1001},\rho_{0110},$ and $\rho_{0101}$ can be represented as following 
\begin{eqnarray}\label{even432}
\nonumber
    \rho_{1010}&=&\frac{1}{4}(\mathrm{I}+A_{4,2}+A^{\prime}_{4,2}+\tilde{A}_{4,2}); \ \ \ 
    \rho_{1001}=\frac{1}{4}(\mathrm{I}+A_{4,2}-A^{\prime}_{4,2}-\tilde{A}_{4,2})\\
    \rho_{0110}&=&\frac{1}{4}(\mathrm{I}-A_{4,2}+A^{\prime}_{4,2}-\tilde{A}_{4,2}); \ \ \ 
    \rho_{0101}=\frac{1}{4}(\mathrm{I}-A_{4,2}-A^{\prime}_{4,2}+\tilde{A}_{4,2})
\end{eqnarray}

Consequently, the remaining eight quantum states can be represented in terms of the two commuting set $[A_{4,3},A^{\prime}_{4,3}]$ and $[A_{4,4},A^{\prime}_{4,4}]$ as
\begin{eqnarray}\label{even433}
\nonumber
    \rho_{0001}&=&\frac{1}{4}(\mathrm{I}+A_{4,3}+A^{\prime}_{4,3}+\tilde{A}_{4,3});\ \ \ \ \  \ \ \ \rho_{0100}=\frac{1}{4}(\mathrm{I}+A_{4,4}+A^{\prime}_{4,4}+\tilde{A}_{4,4})\\
    \nonumber
    \rho_{1101}&=&\frac{1}{4}(\mathrm{I}+A_{4,3}-A^{\prime}_{4,3}-\tilde{A}_{4,3});\ \ \ \ \  \ \ \ \rho_{0111}=\frac{1}{4}(\mathrm{I}+A_{4,4}-A^{\prime}_{4,4}-\tilde{A}_{4,4})\\
    \rho_{0010}&=&\frac{1}{4}(\mathrm{I}-A_{4,3}+A^{\prime}_{4,3}-\tilde{A}_{4,3});\ \ \ \ \  \ \ \ \rho_{1000}=\frac{1}{4}(\mathrm{I}-A_{4,4}+A^{\prime}_{4,4}-\tilde{A}_{4,4})\\
    \nonumber
    \rho_{1110}&=&\frac{1}{4}(\mathrm{I}-A_{4,3}-A^{\prime}_{4,3}+\tilde{A}_{4,3});\ \ \ \ \  \ \ \ \rho_{1011}=\frac{1}{4}(\mathrm{I}-A_{4,4}-A^{\prime}_{4,4}+\tilde{A}_{4,4})
\end{eqnarray}
such that the parity information remains intact.

Given the parity-set $\mathbb{G}_{4,3}$, from Eqs. (\ref{even431}) and (\ref{even432}), parity condition for the even parity set of $s=1111$ is calculated as $\sum_{x|s.x=0}\rho_{x}=\mathbb{I} \otimes \mathbb{I}$. Thus, it is easy to show that for parity element $s=1111$: $\sum_{x|s.x=0}\rho_{x}=\sum_{x|s.x=1}\rho_{x}=\mathbb{I} \otimes \mathbb{I}$ i.e., PO condition is naturally satisfied. But for the 
elements $s=1110,1101,1011$, and $0111$, four different nontrivial relations between Alice’s observables can be defined. 

For the parity element $s=1110$ the even parity set is $\{0000,1100,1010,0110,0001,1101,0111,1011\}$ and the odd parity set is $\{0011,1111,1001,0101,0010,1110,0100,1000\}$. Using Eqs. (\ref{even431}), (\ref{even432}) and (\ref{even433}), the observable relation for the even parity set of the element $s=1110$ can be derived as
\begin{equation}
    \rho_{0000}+\rho_{1100}+\rho_{1010}+\rho_{0110}+\rho_{0001}+\rho_{1101}+\rho_{0111}+\rho_{1011}=4\mathrm{I}+A^{\prime}_{4,1}+A^{\prime}_{4,2}+A_{4,3}-A^{\prime}_{4,4}
\end{equation}
Thus, it is straightforward to show that 
\begin{equation}
    s=1110:\ \ \ \ \sum_{x|s.x=0}\rho_{x}=\sum_{x|s.x=1}\rho_{x}\implies A^{\prime}_{4,1}+A^{\prime}_{4,2}+A_{4,3}-A^{\prime}_{4,4}=0
\end{equation}
Similarly, from Eqs. (\ref{even431}), (\ref{even432}) and (\ref{even433})
the other three PO conditions are derived as
\begin{eqnarray}
\nonumber
	&&s=1101:\ \ \ \ \ A_{1}-A_{4,2}-A_{4,3}^{\prime}-A_{4,4}=0;\\
	\nonumber
	&&s=1011:\ \ \ \ \ A_{4,1}^{\prime}-A_{4,2}^{\prime}-A_{4,3}-A_{4,4}^{\prime}=0; \\
	&&s=0111:\ \ \ \ \ A_{4,1}+A_{4,2}-A_{4,3}^{\prime}+A_{4,4}=0.
\end{eqnarray}

The above four nontrivial parity conditions provide the constraint relations 
\begin{equation}\label{ntpo43}
    A_{4,1}=A_{4,3}^{\prime},\  \  \  \ \text{and}\  \  \  \  \  A_{4,1}^{\prime}=A_{4,4}^{\prime}
\end{equation}
which has to be satisfied in a PO task.

Thus, imposing the PO relations from Eq. (\ref{ntpo43}), the quantum success probability can then be casted as
\begin{eqnarray}\label{de}
	(\mathcal{P}_4^{g_4=3})_{Q} =\dfrac{1}{2}+\dfrac{\mathrm{tr}[\Delta_{4}^{g_4=3}]_Q}{128}
\end{eqnarray} 
where the correlation function $\Delta_{4}^{g_4=3}$ is  
\begin{eqnarray}\label{d13}
	\nonumber
	\Delta_{4}^{g_4=3} &=& (A_{4,1}+A_{4,2}+A_{4,3}+A_{4,4}) B_{4,1}+ (A_{4,1}-A_{4,2}+A_{4,3}-A_{4,4}) B_{4,2}\\
	&+&
	 (2A'_{4,1}+A'_{4,2}+A_{4,1}) B_{4,3}+ (2A'_{4,1}-A'_{4,2}-A_{4,1}) B_{4,4}.
\end{eqnarray} 
The expression $\Delta_{4}^{g_4=3}$ has the preparation noncontextual bound $32$ which consequently provides the  preparation noncontextual success probability $(\mathcal{P}_4^{g_4=3})_{pnc}=3/4$.

To find the quantum upper bound of $\mathrm{tr}[\Delta_{4}^{g_4=3}]_{Q}$, we define 

\begin{eqnarray}
\label{sos1}
\nonumber
	B_{4,1}=\frac{(A_{4,1}+A_{4,2}+A_{4,3}+A_{4,4})}{\omega_{4,1}};\ \ \
	B_{4,3}=\frac{(2A'_{4,1}+A'_{4,2}+A_{4,1})}{\omega_{4,3}} \\
		B_{4,2} =\frac{(A_{4,1}-A_{4,2}+A_{4,3}-A_{4,4})}{\omega_{4,2}} ;\ \ \
	B_{4,4}=\frac{(2A'_{4,1}-A'_{4,2}-A_{4,1})}{\omega_{4,4}} 
\end{eqnarray}
where $\omega_{4,i}$'s are suitable positive number which can be explicitly written as
\begin{eqnarray}
\label{soso}
\nonumber
	\omega_{4,1}&=&||(A_{4,1}+A_{4,2}+A_{4,3}+A_{4,4}) ||;\ \ \ 
	\omega_{4,3}=||(2A'_{4,1}+A'_{4,2}+A_{4,1}) ||\\
	\omega_{4,2}&=&||(A_{4,1}-A_{4,2}+A_{4,3}-A_{4,4}) ||;\ \ \  
	\omega_{4,4}=||(2A'_{4,1}-A'_{4,2}-A_{4,1}) ||.
\end{eqnarray}
where $||\mathcal{O}||=\sqrt{\lambda_{max}\Big(\mathcal{O}^{*}\mathcal{O}\Big)}$ is the trace norm where $\lambda_{max}$ denotes the maximum eigenvalue of $\mathcal{O}$.

Putting Eqs. (\ref{sos1}-\ref{soso}) in Eq. (\ref{d13}), we find
\begin{equation}
	\label{max}
max\left(\mathrm{tr}[\Delta_{4}^{g_4=3}]\right)=4 \ max(\sum_{i=1}^{4}\omega_{4,i})
\end{equation}

where $\omega_{4,i}$'s can be written as
\begin{eqnarray}
	\nonumber
	\omega_{4,1}&=&\sqrt{\lambda_{max}\Big(\  4\mathbb{I}+\{A_{4,1},A_{4,2}\}+\{A_{4,1},A_{4,3}\}+\{A_{4,1},A_{4,4}\}+\{A_{4,2},A_{4,3}\}+\{A_{4,2},A_{4,4}\}+\{A_{4,3},A_{4,4}\} \Big)},\\
	\nonumber
	\omega_{4,2}&=&\sqrt{\lambda_{max}\Big(\ 4\mathbb{I}-\{A_{4,1},A_{4,2}\}+\{A_{4,1},A_{4,3}\}-\{A_{4,1},A_{4,4}\}-\{A_{4,2},A_{4,3}\}+\{A_{4,2},A_{4,4}\}-\{A_{4,3},A_{4,4}\} \Big)},\\
	\nonumber
	\omega_{4,3}&=&\sqrt{\lambda_{max}\Big(\ 6\mathbb{I}+2\{A_{4,1},A_{4,1}^{\prime}\}+2\{A_{4,1},A_{4,2}^{\prime}\}+2\{A_{4,1}^{\prime},A_{4,2}^{\prime}\} \Big)},\\
	\omega_{4,4}&=&\sqrt{\lambda_{max}\Big(\ 6\mathbb{I}-2\{A_{4,1},A_{4,1}^{\prime}\}+2\{A_{4,1},A_{4,2}^{\prime}\}-2\{A_{4,1}^{\prime},A_{4,2}^{\prime}\} \Big)}.
\end{eqnarray}    

The maximum value of $\left(\mathrm{tr}[\Delta_{4}^{g_4=3}]_{Q}\right)$ in Eq. (\ref{max}) has to be derived by considering PO condition  along with the condition $[A_{4,i},A_{4,i}^{\prime}]=0$ with $i\in \{1,...,4\}$.  For a two-qubit system, it is straightforward to find $max\left(\mathrm{tr}[\Delta_{4}^{g_4=3}]\right)=4(\sqrt{10}+5\sqrt{2})$. Consequently, by using Eq. (\ref{de}), we find the optimal success probability  $(\mathcal{P}_4^{g_4=3})_{Q} \approx 0.819>(\mathcal{P}_4^{g_4=3})_{pnc}=0.75$. 

A choice of Alice's observables for which the above quantum bound can be obtained is the following. 
\ba
&&A_{4,1} =A_{4,3}^{\prime}= \sigma_{x} \otimes \sigma_{x} ,\ \  A_{4,1}^{\prime} = A_{4,4}^{\prime}=\sigma_{y} \otimes \sigma_{y} ;\\
\nonumber
&&A_{4,2}= \sigma_{x} \otimes \sigma_{y} , A_{4,2}^{\prime}= \sigma_{y} \otimes \sigma_{x} ; A_{4,3} = \sigma_{z} \otimes \sigma_{y} ; A_{4,4} = \sigma_{z} \otimes \sigma_{x}\ea

\section{3-bit entanglement-assisted PORAC}\label{app3ea}

Our proposed prepare-and-measure scenario can also be converted into  entanglement-assisted scenario. In such a case, Alice and Bob share an entangled state $\rho_{AB} \in \mathbb{C}^4\otimes \mathbb{C}^4$. Alice encodes eight input states by performing sequential projective measurement randomly on two pairs of commuting observables ($A_{3,i}$ and $A^{\prime}_{3,i}$) with $i=1,2$ on her system as depicted in \figurename{ 1}. Alice perform a non-selective degeneracy breaking measurement of an observable (say, $A^{\prime}_{3,i}$) before the measurement of $A_{3,i}$ to encode the bits in rank-1 density matrices.  The steered (unnormalized) states  $\tilde{\rho}_{x}$ can be compactly written as
\begin{eqnarray}
\tilde{\rho}_{x}\equiv\ket{\psi_{x}}\bra{\psi_{x}}=\mathrm{tr_A}[(P_{A_{3,i}}^{a} \otimes \mathbb{I})(P_{A^{\prime}_{3,i}}^{a^{\prime}}\otimes \mathbb{I})\hspace{.1cm}\rho_{AB}]
\end{eqnarray} 
where $a, a^{\prime}=\{+,-\}$ and $\rho_{x}=\tilde{\rho}_{x}/Tr[\tilde{\rho}_{x}]$.

\begin{figure}[ht]
	\label{Fig}
	\includegraphics[width=0.5\linewidth]{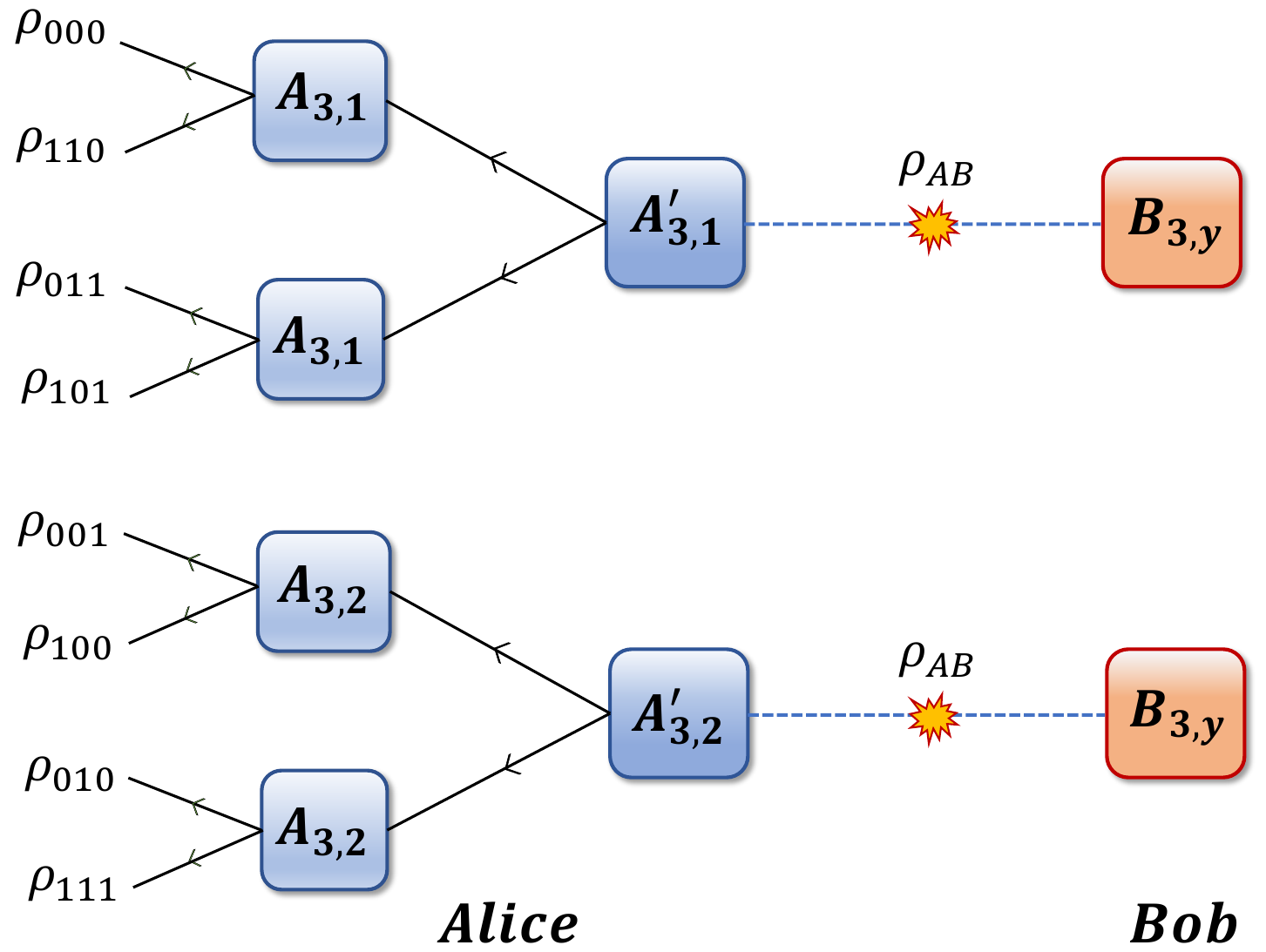}
\caption{The encoding scheme for  PORAC $\mathbb{G}_{3,3}$ by using degeneracy breaking non-selective sequential measurements of commuting observables}
\end{figure} 

Note here that given the parity element $s=\{111\}$, in quantum theory the PO condition provides 
\begin{equation}
    1/4(\rho_{000}+\rho_{011}+\rho_{110}+\rho_{101})=1/4(\rho_{001}+\rho_{010}+\rho_{100}+\rho_{111})=(\mathbb{I\otimes I})/4,
\end{equation} i.e, the mixture of the four states corresponding to the even parity is identical to the mixture of the four states corresponding to odd parity. Thus, the parity-oblivious condition is naturally satisfied for such encoding.

In a two-qubit system at most three mutually commuting observables are available and hence product of any two will return the third one.  Moreover, for a two-qubit system the choice of basis for doubly degenerate eigen-subspace are not unique and can be written in infinitely many ways. Thus, to optimize the quantum success probability, in Alice's encoding scheme we considered the prime observables to be same (i.e., $A^{\prime}_{3,1}=A^{\prime}_{3,2}=A^{\prime}_{3}$) for different set of non-selective degeneracy breaking measurement corresponding to $A_{3,1}$ and $A_{3,2}$ respectively. Such a encoding scheme not only satisfies the restriction in quantum theory as given in Eq. (\ref{constraint}) in the main text but can also be equivalently represented in a preparation noncontextual ontological model of the operational quantum theory. Interestingly, this restriction in a preparation noncontextual model further reproduces the preparation noncontextual bound for a $3$-bit PORAC.  

 On the other hand, Bob performs three projective measurements $ \{\Pi_{3,y}^{0},\Pi_{3,y}^{1}\}$ with $y=1,2,3$, with $b= x_y= 0,1$.	Then, using Eq. (\ref{QSP}) from the main text, a similar derivation will give the quantum success probability in terms of Bell expression as

\begin{equation}
\label{}
(\mathcal{P}_{3}^{g_3=3})_Q=\dfrac{1}{2}+\dfrac{\langle\mathcal{B}_{3}^{g_3=3}\rangle_Q}{12}
\end{equation} where the Bell expression $\mathcal{B}_{3}^{g_3=3}$  can be written as
\begin{eqnarray}
\label{eabell}
\mathcal{B}_{3}^{g_3=3}=(A_{3,1}+A_{3,2})\otimes B_{3,1}+({\tilde A}_{3,1}-{\tilde A}_{3,2})\otimes B_{3,3}+ 2A^{\prime}_{3}\otimes B_{3,2}. 
\end{eqnarray}

The optimal quantum value of the Bell expression  in Eq. (\ref{eabell}) is derived as \ba 
(\mathcal{B}_{3}^{g_3=3})_{Q}^{opt} = 2+2\sqrt{2}.\ea 
when $B_{3,1} = (A_{3,1} +A_{3,2})/\sqrt{2}$ , $B_{3,2} = A^{\prime}_{3}$ and $B_{3,3} = (\tilde{A}_{3,1} -{\tilde A}_{3, 2})/\sqrt{2}$, and the shared entangled state is $
\ket{\psi_{AB}}=\ket{\phi_{AB}^{+}}^{\otimes 2}$ where 
\ba
\ket{\phi_{AB}^{+}}=\frac{\ket{00}+\ket{11}}{\sqrt{2}}
\ea	
 Thus, the optimal quantum success probability is
\begin{equation}	\label{eaopta}(\mathcal{P}_{3}^{g_3=3})_{Q}^{opt}=\dfrac{1}{2}+\dfrac{1+\sqrt{2}}{6}\approx0.902
\end{equation} 
which remains same as in the prepare-and-measure scenario. 

Let us provide a simple strategy which saturates the optimal quantum success probability  in Eq. (\ref{eaopta}). In entanglement-assisted scenario Alice is allowed to send two bits  information to Bob. Upon receiving the bits from Alice, Bob performs projective measurements based on a prior strategy. Suppose Alice sends $\{00,01,10,11\}$ to Bob. Note that, $00$ has the equal probability to be chosen from $s$-parity $0$ and $s$-parity $1$, and same for other cases. Hence, Alice's communication does not reveal the parity information to Bob. Let, Alice always sends the first two bits. If Alice communicates $00$, Bob  performs the projective measurements $\Pi_{3,1}^{0}, \Pi_{3,2}^{0},\Pi_{3,3}^{0}$ to learn about first, second and third bit respectively. Similarly,  if Alice communicates $01$, Bob  performs $\Pi_{3,1}^{0}, \Pi_{3,2}^{1},\Pi_{3,3}^{1}$ on his system. Again, for Alice's communication of $10$ and $11$, Bob  performs  projective measurement  $\Pi_{3,1}^{1}, \Pi_{3,2}^{0},\Pi_{3,3}^{1}$ and  $\Pi_{3,1}^{1}, \Pi_{3,2}^{1},\Pi_{3,3}^{0}$ respectively. This strategy provides the optimal quantum success probability given in Eq. (\ref{eaopta}).

Finally, we note that we have used a pair of two-qubit entangled states here but this protocol works for a pair of entangled states in any arbitrary dimension.

 \section{Proposed experimental scheme for 3-bit PORAC}\label{experimet}
\begin{figure}[ht]
	\centering
	\label{Fig}
	\includegraphics[width=0.7\linewidth]{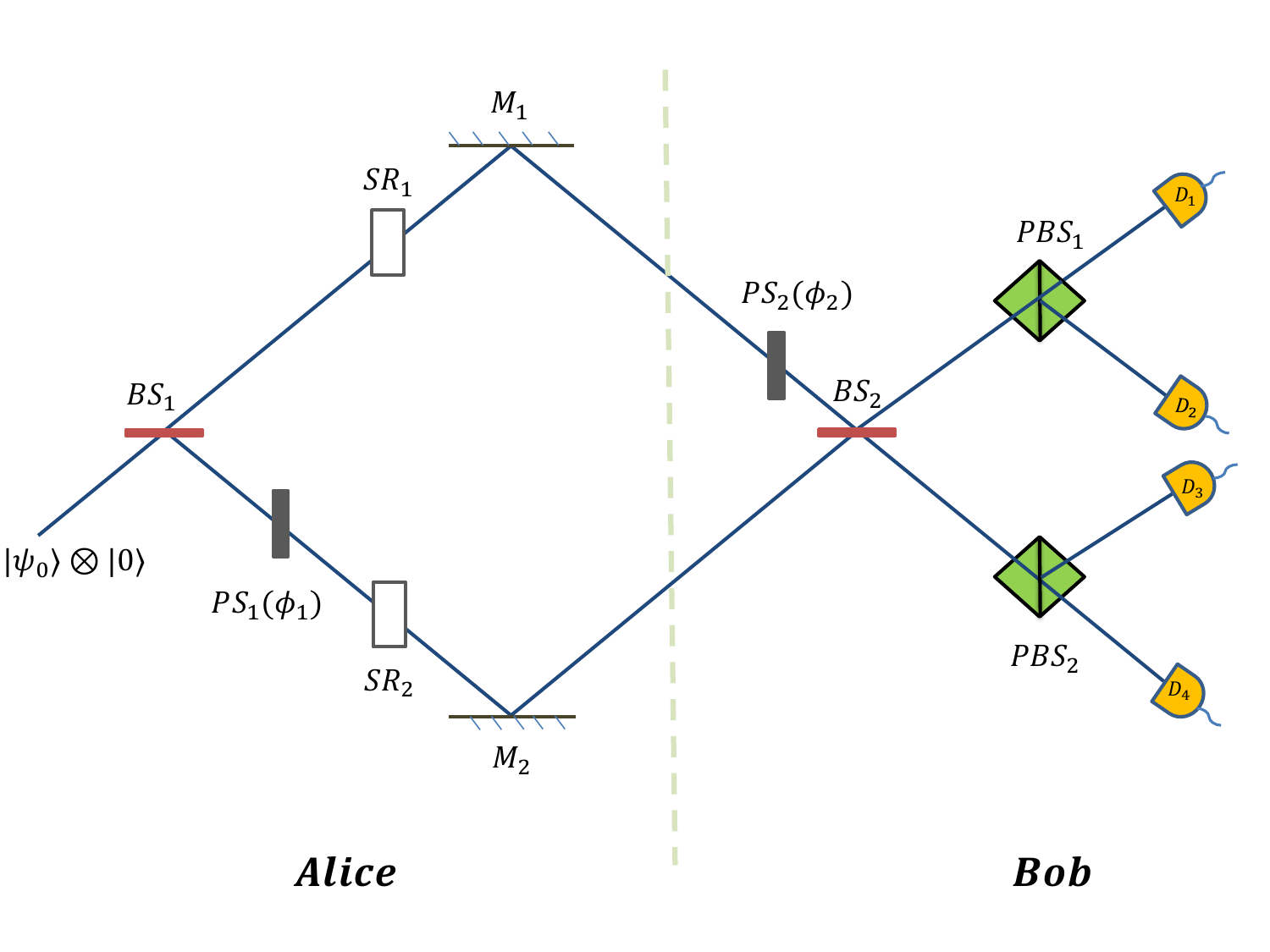}
	\caption{A schematic of the experimental setup for the PORAC $\mathbb{G}_{3,3}$ in prepare-and-measure scenario }
\end{figure} 
We provide a sketch of the experimental setup corresponding to the parity-set $\mathbb{G}_{3,3}$ in the prepare-and-measure scenario by using a suitable Mach-Zehender setup. For $\mathbb{G}_{3,3}$, the eight pure states for Alice's encoding are the following; $\rho_{000}=|+ 0\rangle \langle0 +|$, $\rho_{110}=|+1\rangle \langle1 +|$, $\rho_{011}=|-0\rangle \langle 0 -|$ and $\rho_{101}=|-1\rangle\langle 1-|$ are the common eigenstates of $A_{3,1} = \sigma_{x} \otimes \sigma_{x}$ , ${\tilde A}_{3,1} = \mathbb{I} \otimes \sigma_{x}$ and  $A_{3}^{\prime} = \sigma_{x} \otimes \mathbb{I}$ are the $s$-parity $0$ set. The $s$-parity $1$ set are given by $\rho_{001}=|++\rangle \langle ++|$, $\rho_{100}=|-+\rangle \langle +-|$, $\rho_{010}=|+-\rangle \langle -+|$ and $\rho_{111}=|--\rangle \langle --|$ which are the common eigenstates of  $A_{3,2}= \sigma_{x} \otimes \sigma_{z}$, $A^{\prime}_{3} = \sigma_{x} \otimes \mathbb{I}$ and ${\tilde A}_{3,2} = \mathbb{I} \otimes \sigma_{z}$.

The Mach-Zehender setup shared by Alice and Bob is depicted in \figurename{ 2}. The entire setup consists of a $50 : 50$ beam splitter ($BS_1$), an asymmetric beam-splitter $(BS_{2}$), two polarizing beam splitters ($PBS_{1}$ and $PBS_{2}$), two polarization(spin)-rotators ($SR_1$ and $SR_2$), two mirrors ($M_{1}$ and $M_{2}$),  and two phase-shifters ($PS_1$ and $PS_{2}$). Upon receiving the joint path-spin state $|\psi_{0}\rangle\otimes |0\rangle$ of a single particle, by using $BS_{1}$, $SR_1$, $SR_2$ and $PS_{1}$, Alice prepares the following general state 
\begin{equation}
	\label{IS2}
	\ket{\psi_{x}}=\dfrac{(\ket{\psi_{1}}+ie^{i\phi_{1}}\ket{\psi_{2}})}{\sqrt{2}} \otimes(cos\theta\ket{0}+sin\theta\ket{1})
\end{equation}
where $\ket{\psi_{1}}\equiv \ket{0}$ and $\ket{\psi_{2}}\equiv \ket{1}$ are the path states. Two different values of $\phi_{1} \ (\pi/2$ and $3\pi/2$), and four different values of $\theta \ (0,\pi/2,\pi/4$ and $3\pi/4)$ provide the required eight input pure states. 

Bob decodes the information  by using three measurements $B_{3,y}=B_{3,p}\otimes\sigma_{3,\theta}$ implemented by using $PS_{2}$, $BS_{2}$, $PBS_{1}$ and $PBS_{2}$. Here $B_{3,p}$ and $\sigma_{3,\theta}$ are the relevant path and polarization observables respectively. He implements the relevant path measurements by using the $PS_{2}$ and $BS_{2}$ which can be written as $B_{3,p}=P(\psi_{3})-P(\psi_{4})$, where the eigenvalues $\pm 1$ of $B_{3,p}$ pertain to measurement in either $\ket{\psi_{3}}$ or $\ket{\psi_{4}}$ respectively which are unitarily related to the states $\ket{\psi_{1}}$ and $\ket{\psi_{2}}$ as 
\begin{eqnarray}
	\label{OP}
	\ket{\psi_{3}}=i\alpha e^{i\phi_{2}}\ket{\psi_{1}}+\beta\ket{\psi_{2}}, \  \ket{\psi_{4}}=\beta e^{i\phi_{2}}\ket{\psi_{1}} +i\alpha\ket{\psi_{2}}
\end{eqnarray}
where $\alpha (\beta)$ are reflectivity (tranmitivity) satisfying $|\alpha|^2+|\beta|^2=1$. Taking, $\ket{\psi_{1}}=\ket{0}$ and  $\ket{\psi_{2}}= \ket{1}$ ,
and using the relations in Eq. (\ref{OP}) Bob's path observables $B_{3,p}$ can can be represented as
\begin{equation}
\begin{pmatrix}
	\alpha^2-\beta^2 & 2i\alpha\beta e^{i\phi_{2}} \\
	-2i\alpha\beta e^{-i\phi_{2}} & \beta^2 -\alpha^2
\end{pmatrix}.
\end{equation}
Different values of $\alpha(\beta)$ and $\phi_{2}$ will lead to different path observables and the polarization observables can be implemented by $PBS_1$ and $PBS_{2}$. This provides the required observables  $B_{3,y}$ for Bob with $y=1,2,3$. 
\end{widetext}

			\section*{Aknowledgement} PR acknowledges the support from the project
			DST/ICPS/QuST/Theme-2/2019/General Project Q-90. AKP acknowledges the support from the research grant MTR/2021/000908.

		\end{document}